\documentclass{article}

\usepackage{PRIMEarxiv}

\usepackage[utf8]{inputenc} 
\usepackage[T1]{fontenc}    
\usepackage{hyperref}       
\usepackage{url}            
\usepackage{booktabs}       
\usepackage{amsfonts}       
\usepackage{nicefrac}       
\usepackage{microtype}      
\usepackage{lipsum}
\usepackage{fancyhdr}       
\usepackage{graphicx}       
\graphicspath{{media/}}     
\usepackage{amssymb}
\usepackage{pifont}
\usepackage{relsize}
\usepackage{color, colortbl}
\usepackage{xcolor}   
\usepackage{booktabs}
\usepackage{pifont}
\usepackage{rotating}
\usepackage{array}
\usepackage{arydshln}       
\usepackage{multicol}
\usepackage{tablefootnote}
\usepackage{tabularx}
\usepackage{makecell}
\usepackage{float}
\usepackage{caption}
\usepackage[para]{threeparttable}
\usepackage{csquotes,booktabs,tabularx,ragged2e}

\newcommand{\xmark}{\ding{55}}
\newcommand{\cmark}{\checkmark}
\usepackage{lscape}
\definecolor{CHANNEL}{HTML}{D5C2FF}
\definecolor{KERNEL}{HTML}{FFFCBD}
\definecolor{ORANGE}{HTML}{FFC3DE}
\definecolor{HYBRID}{HTML}{C3DFCA}

\definecolor{teal}{RGB}{0, 128, 128}
\definecolor{orange}{RGB}{255, 140, 0}
\definecolor{lightpurple}{RGB}{173, 216, 230}
\definecolor{green}{RGB}{0, 128, 0}

\definecolor{lightblue}{RGB}{173, 216, 230}
\definecolor{lightorange}{RGB}{255, 192, 128}
\definecolor{lightgreen}{RGB}{144, 240, 144}
\definecolor{lightpink}{RGB}{255, 182, 193}

\title{Mixing Solutions in Bitcoin and Ethereum Ecosystems: A Review and Tutorial }

\author{
  Alireza Arbabi\thanks{These authors contributed equally to this work.} \\
  Department of Electrical and Computer Engineering\\
  University of Tehran \\
  Tehran, Iran \\
  \texttt{alireza.arbabi@ut.ac.ir} \\
\And
  Ardeshir Shojaeinasab$^*$ \\
  Department of Electrical and Computer Engineering\\
  University of Victoria \\
  Victoria, Canada \\
  \texttt{ardeshir@uvic.ca} \\
\And
  Behnam Bahrak \\
  Tehran Institute for Advanced Studies \\
  Tehran, Iran \\
  \texttt{b.bahrak@teias.institute} \\
\And
  Homayoun Najjaran\thanks{Corresponding author} \\
  Department of Mechanical Engineering\\
  University of Victoria \\
  Victoria, Canada \\
  \texttt{najjaran@uvic.ca}
}

\begin{document}

\maketitle

\begin{abstract}
This manuscript presents an exhaustive review of blockchain-based mixing services, aiming to fill the existing gap between academic innovations and real-world implementations. Starting with an identification of the core functionalities and techniques employed by mixing services, the paper delves into detailed explanations of these operational mechanisms. It further outlines an evaluation framework tailored for a rigorous assessment, highlighting the key vulnerabilities and strengths of various solutions. In addition, the study identifies potential attack vectors that compromise these services. The paper explores the dual nature of mixing services: while they contribute to the preservation of privacy—a cornerstone of blockchain technologies—they can also facilitate illicit activities. By addressing key research questions, this study not only offers a comprehensive overview of the current state of mixing services but also sets the stage for future academic discourse in this evolving field.

\end{abstract}


\section{Introduction}

The advent of blockchain technology has had a significant impact on the financial landscape by introducing cryptocurrencies as a new, decentralized form of digital assets \cite{nakamoto2008bitcoin}. While these digital currencies promise a level of privacy and security, the fundamental structure of blockchain technology - based on transparent and immutable ledgers - presents challenges to individual privacy \cite{sompolinsky2015secure}. Given the publicly accessible full transaction record, monitoring the movement of funds between cryptocurrency addresses is possible. Consequently, if a user receives funds from an identified suspicious address, both the user's address and funds will be flagged as questionable, leading to a reduction in the user's privacy and anonymity.

The effort of balancing transactional transparency with privacy has led to the development of cryptocurrency mixing services, designed to obscure the source and destination of transactions. These mixing services occupy a controversial status. On one hand, they serve as vital tools for privacy preservation and equity. On the other hand, they hold the potential for misuse in money laundering, sanction evasion, ransomware money transfer, and other illicit activities. For instance, studies have demonstrated that SilkRoad extensively utilized crypto mixers to obfuscate its users' funds \cite{christin2013silkRoad}, and various ransomware like Wannacry employed mixing solutions to conceal the flow of their funds \cite{bistarelli2018wannacry}. Additionally, the Lazarus Group has leveraged mixing services to obscure over \$991 million of stolen funds on behalf of North Korea \cite{CypherTrace2023,chainalysisCrimeReport}.

Considering this landscape, this review paper provides a comprehensive study of blockchain-based mixing services, targeting academic solutions as well as real-world practitioners. Given the growing influence of mixing services and the profound ethical and technical dilemmas they present, there is an urgent need for a comprehensive examination that goes beyond disciplinary boundaries. This review aims to fill this gap by not only examining existing models and solutions but also by proposing a unified framework for their evaluation. This synthesis is essential to accelerate research efforts, guide policy-making, and assist in the development of more robust and secure mixing services. To achieve these objectives, we specify and aim to answer the following critical research questions:

\begin{itemize}
    \item What are the most critical mixing techniques in cryptocurrencies?
    \item What are the main mixing services provided in academia, and how do they work?
    \item What are the available real-world solutions, and how do they operate?
    \item How closely do real-world and academic mixing solutions align?
    \item Can we propose a framework for evaluating mixing services, and how should these be evaluated?
    \item To what extent do mixing services render transactions fully anonymous, and how secure are they?
    \item What open challenges exist in this field that demand scholarly attention?
\end{itemize}

Our investigations involve a survey of academic frameworks alongside an analysis of existing market solutions by forensics of current attacks performed on each service to convey some information from the dark side of target mixing services. The paper proposes a first-of-its-kind set of evaluation criteria for mixing services and provides a detailed analysis on what are the weaknesses and strengths of different mixing solutions and the potential for improvement case by case. 
The paper also provides an exploration of potential vulnerabilities and outlines a roadmap for future research and development in the field. Ultimately, the research aims to contribute both to the academic discourse and to real-world implementations by identifying ongoing challenges such as fund traceability in a full path from sender to the receivers and the unknown territory of cross-chain mixing.

\section{Background}

The introduction of mixers marked a significant transformation in cryptocurrencies, directly addressing the challenges of trust, security, privacy, and efficiency. Essential to the operation and growing acceptance of these blockchain-based mixers are the foundational technologies and principles. We will provide a brief overview of some of these key concepts below before going over how the mixers operate in detail:

\paragraph{Blockchain:} Introduced in 2008 by the pseudonymous Satoshi Nakamoto for the cryptocurrency Bitcoin, a blockchain is a decentralized and distributed digital ledger technology used to record transactions across multiple computers in a way that ensures the data can only be modified once it's been recorded. Once a block of data has been added to the blockchain, it becomes virtually immutable, protected from alteration without altering all subsequent blocks and the consensus of the network. This characteristic ensures data integrity and transparency. Each block in the chain contains a number of transactions, and every time a new transaction occurs on the blockchain, a record of that transaction is added to every participant's ledger. This decentralized nature of the system ensures that no single entity has control over the entire blockchain, and all transactions are publicly recorded, ensuring transparency and trustworthiness ~\cite {Gaikwad2020Overview}.

\paragraph{Transactions:} At the heart of Bitcoin or any other cryptocurrency functioning is its transaction-based public ledger. Transactions here are a play of Unspent Transaction Outputs (UTXOs), which are used wholly in transactions. Given the improbability of a UTXO matching an exact spending amount, most Bitcoin transactions result in two outputs. One is received by the intended recipient, while the change is sent back to the sender at a new address. Additionally, transaction metadata encompasses public keys, UTXOs, transaction size, and its unique hash. With inputs signed using the sender's private key, validity is readily verifiable via the sender's public key ~\cite{nakamoto2008bitcoin}.

\paragraph{Consensus Mechanisms:} One of the revolutionary aspects of blockchain technology is the elimination of central authority, ensuring transactions' validation through consensus mechanisms. These algorithms ensure all nodes in the network agree upon the truth. Bitcoin, for example, utilizes the Proof-of-Work (PoW) mechanism, where participants (miners) solve cryptographic puzzles to validate transactions and add new blocks. However, concerns regarding energy consumption and scalability have led to the exploration of alternative consensus mechanisms. Proof-of-Stake (PoS) is one such alternative where validators are chosen to create new blocks based on the number of coins they hold and are willing to "stake" as collateral. Each consensus mechanism has its trade-offs in terms of security, decentralization, and efficiency \cite{nakamoto2008bitcoin, KingNadal2012}.
    
\paragraph{Cryptographic Hash Functions:} A cornerstone of blockchain technology, cryptographic hash functions, ensure data integrity and security. These functions take an input and produce a fixed-size string of bytes, typically a digest that is unique to each unique input. The SHA-256, predominantly used in Bitcoin, is one such algorithm. These functions are crucial for generating public and private keys, forming blocks, and maintaining data consistency and integrity across the decentralized network \cite{NIST2012}.

\paragraph{Privacy in Cryptocurrencies:} While cryptocurrencies offer a robust mechanism for transaction security, privacy remains a nuanced challenge. Traditional banking systems provide privacy by restricting access to transactional information. Cryptocurrencies, on the other hand, announce all transactions publicly. However, privacy is attempted by keeping public keys anonymous, akin to stock exchanges. A key challenge arises when key owners are revealed, potentially unveiling other linked transactions~\cite{nakamoto2008bitcoin}. Furthermore, users' identities can be deanonymized by associating pseudonyms with IP addresses or by abusing Bitcoin's anti-DoS measures ~\cite{Biryukov2014Deanonymisation}.

\paragraph{Smart Contracts:} Beyond traditional transactions, the blockchain ecosystem has given rise to smart contracts. These are digital protocols designed to autonomously execute tasks when predefined conditions are met. Such contracts eliminate third-party interventions and can engender automatic payments, quality controls, and establish trust amongst stakeholders ~\cite{Kumar2019Combating}. Not limited to Bitcoin, these contracts have been enhanced on other platforms due to the introduction of features like Turing-completeness and blockchain-awareness ~\cite{buterin2014next}.
    
\paragraph{Decentralized Finance (DeFi):} DeFi represents a conglomerate of decentralized applications aiming to recreate or enhance traditional financial systems (like lending, borrowing, and derivatives) without intermediaries using blockchain technology. Predominantly built upon Ethereum, these platforms leverage smart contracts to ensure transparency, openness, and global accessibility. As DeFi platforms grow, they offer the potential for a more inclusive financial system and present challenges and complexities related to security and regulatory oversight \cite{Schar2021}.

\paragraph{Layer-2 Solutions:} As blockchain networks like Bitcoin and Ethereum became more popular, they faced scalability issues, with transaction speeds being a primary concern. Layer-2 solutions are protocols built on top of a primary blockchain (Layer-1) to increase the transaction throughput. One notable example is Bitcoin's Lightning Network, which facilitates off-chain transactions by opening bilateral channels between parties, thus providing faster and cheaper transactions. Similarly, Ethereum is exploring various Layer-2 solutions, like rollups, to alleviate its scalability constraints \cite{PoonDryja2016}.
    
\paragraph{Zero Knowledge Proof (ZKP):} An intriguing cryptographic tool, ZKP allows for claim verification by an individual without revealing any supportive information. This concept, which emerged in the 1980s, was pioneered by Goldwasser, Micali, and Rackoff from MIT~\cite{fiege1987zero}.

In summation, the landscape of cryptocurrencies is underpinned by a melange of technical innovations and cryptographic techniques, which collectively aim to create a secure, transparent, and efficient digital monetary ecosystem.

\section{Cryptocurrency Mixers}\label{sec:3-functionalities}
In order to analyze how mixing services provide anonymity for their users, we will first explain the definition of mixing. Then, we will take a detailed look at the various obfuscation methods and techniques used by academic and real-world services. Additionally, we will evaluate these methods based on their assessed functionalities.

\subsection{Mixing Definition}
\begin{figure}[tp]
    \centering
    \includegraphics[scale=0.6]{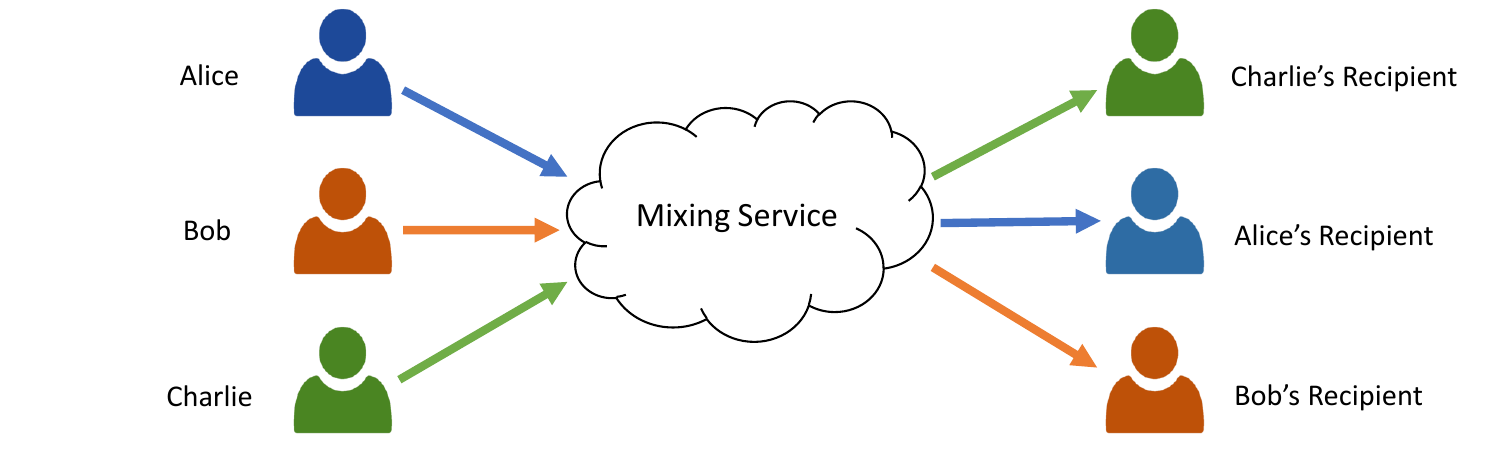}
    \caption{High-level schema of the mixing process. Participants send their funds into the mixing service, and then the service mixes funds and sends them to the specified recipients such that the linking between the corresponding input and outputs is obfuscated.}
    \label{fig:MixingSchema}
\end{figure}

The mixing process in cryptocurrencies, often referred to as coin mixing or coin tumbling, is a technique employed to enhance the privacy and anonymity of transactions within blockchain networks. It involves the pooling of multiple cryptocurrency transactions from various sources and then redistributing them in a manner that obscures their origin and destination. The process typically operates through specialized mixing services or protocols that amalgamate the funds, making it challenging to trace specific coins back to their original owners. By breaking the transactional linkage between addresses and transactions, coin mixing provides a degree of confidentiality, ensuring that the flow of funds remains private and reduces the ability to associate specific transactions with identifiable individuals or entities. This process plays a crucial role in preserving the privacy and anonymity of cryptocurrencies in an increasingly surveilled financial landscape. A high-level schema of the mixing process is depicted in Figure \ref{fig:MixingSchema}.\\

Before discussing the mixing solutions, it's important to introduce two key terms that will be used later: Anonymity Set and Taint Analysis. Anonymity Set refers to the number of participants involved in the mixing process, and Taint refers to the amount of cryptocurrency in an account that came from another account. Taint rates between different addresses can be extracted by tracing back from an address to the origins of its received funds. Taint analysis involves traversing the transaction graph to identify possible connections and relations among different graph vertices (cryptocurrency addresses). This type of analysis helps to deanonymize the activities and connections behind the blockchain.

\subsection{Obfuscating Techniques}
The objective of this subsection is to present a comprehensive overview of obfuscating techniques utilized by mixing services for both Turing-Complete and non-Turing-complete cryptocurrencies. Subsequently, we will examine how frequently each method is used in both academic solutions and real-world services.

\begin{figure}[tp]
    \centering
    \includegraphics[width=0.8\linewidth]{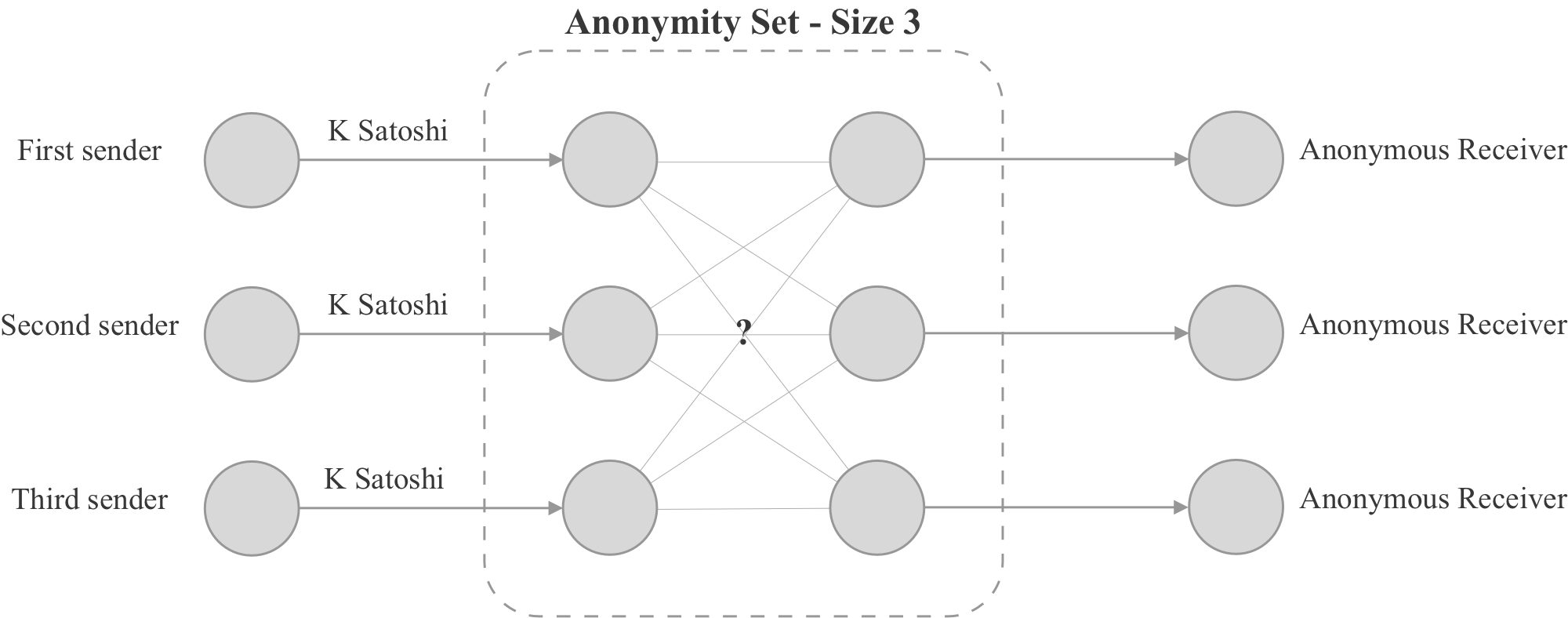}
    \caption{Overview of swapping process \cite{shojaeinasab2022mixing}}
    \label{fig:Swapping}
\end{figure}

\paragraph{Swapping.}
Swapping stands as a prominently employed methodology for obscuring the association between the senders and recipients in a cryptocurrency network \cite{wu2021towards}. The fundamental concept revolves around interchanging the inputs and outputs among diverse participants to keep their relationships secret, as depicted in {Figure}~\ref{fig:Swapping}. By swapping different coins among participants, an intricate meshwork of transactions is created, making it tough to track where the original money came from. The general schema of mixing process is depicted at figure \ref{fig:MixingSchema}

\begin{figure}[t] 
  \centering
  \begin{minipage}{0.45\textwidth}
    \centering
    \includegraphics[width=\linewidth]{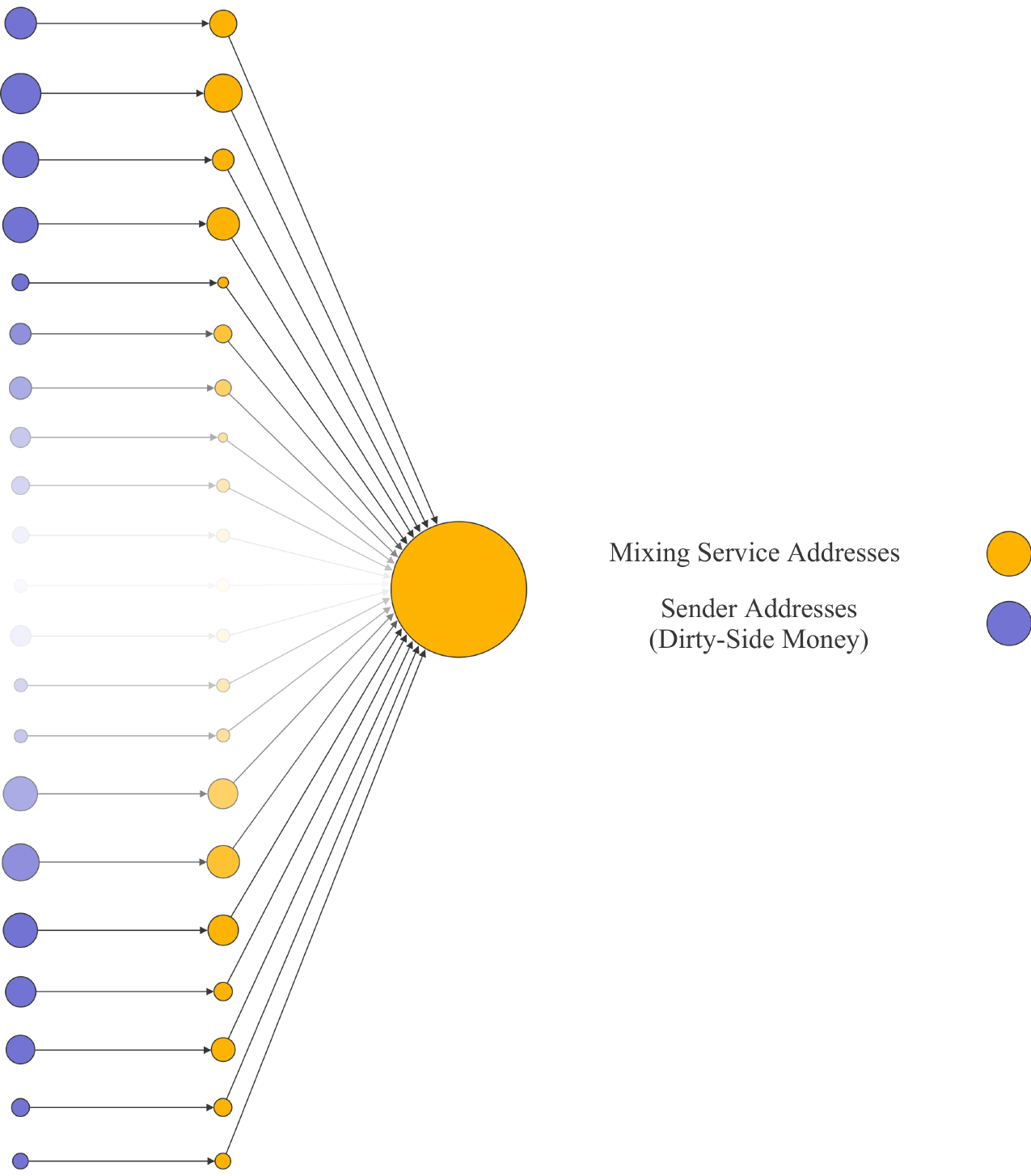}
    \caption{Sample of an Aggregating address \cite{shojaeinasab2022mixing}}
    \label{fig:AggregatingAddress}
  \end{minipage}
  \hfill 
  \begin{minipage}{0.45\textwidth}
    \centering
    \includegraphics[width=\linewidth]{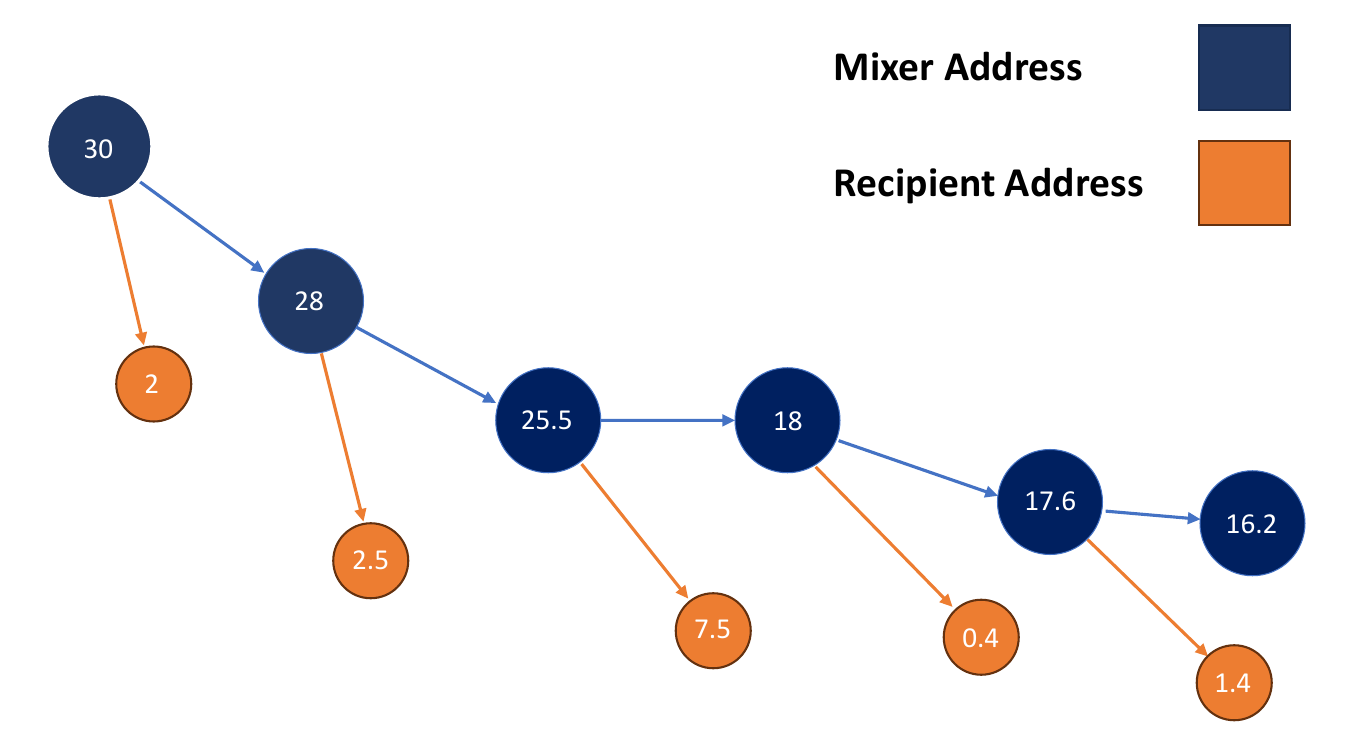}
    \caption{Overview of Peeling Chain schema}
    \label{fig:PeelingChain}
  \end{minipage}
\end{figure}

\paragraph{Aggregating Funds.}
Chang and colleagues discovered specific repetitive transaction patterns in the Bitcoin network, including what they referred to as "sweeper transactions" \cite{chang2018Sweeper}. These transactions, depicted in Figure \ref{fig:AggregatingAddress}, involve a large number of input addresses and one or two output addresses referred to as "Aggregation Addresses". In essence, this method consolidates funds from various addresses into one or two aggregation addresses, resulting in a substantial balance in the newly formed address.  By centralizing all funds into a single input and sending them to the recipient from there, the aggregation address functions as a tool to obscure transactions, which breaks the traceable links between senders and recipients by creating a many-to-one, then one-to-many association.

\paragraph{Peeling Chain.}
A peeling chain is a set of transactions generated by mixing services that form a chain to distribute outputs. The unique property of
the peeling chain is that transactions in the chain are similar to normal user transactions with one input and two outputs \cite{harrigan2016addressClustering}. Therefore, Utilizing peeling chains makes mixing transactions more indistinguishable from a normal user. The schematic of peeling chains is depicted in Figure \ref{fig:PeelingChain}.

\paragraph{Fund Splitting.}
Fund Splitting refers to a process in which users' input coins are divided into smaller denominations and sent through various complex paths, involving multiple participants. Randomization and adding delays in the fund distribution process further complicate the transaction history, making it challenging to link the source and destination of coins.

\paragraph{Chain Hopping.}
Chain hopping is a relatively new and promising technology applied by cryptocurrency mixers to boost privacy and anonymity. The idea behind chain hopping is to add an extra layer of security to the mixing process by utilizing multiple blockchain networks and switching between them. This approach makes it challenging to trace funds, even if the original cryptocurrency source is known.


\paragraph{Randomized Fee.}
In the majority of mixing services, particularly those that are centralized, users are required to pay fees for utilizing the mixing service. Employing a constant fee structure could lead to a detectable usage pattern, potentially compromising the mixing process graph. To counteract this concern, Bonneau et al. \cite{bonneau2014mixcoin} propose the adoption of randomized fees. Implementing a stochastic fee characterized by a continuous range spanning from 0 to a predetermined mixing fee value could eliminate any discernible pattern associated with mixing fees within the mixing transaction graph.

\paragraph{Randomized Delays.}
One of the mixing detection patterns, especially in multi-round and multi-output mixing is looking for time-based patterns in committed transactions to the blockchain. If a mixer consistently uses a fixed time delay in all mixing transactions, mixing patterns can be inferred from this property. To tackle this problem, mixers can put random delays for broadcasting transactions. This way, any predictable timing patterns are avoided, making mixing detection more difficult. 

\paragraph{Third-Party Blinding.}
In centralized mixing setups, a central entity undertakes the mixing process on behalf of participants, affording it comprehensive insight into the connections between senders and recipients. This arrangement raises concerns about potential information leakage, as users must rely on a third party's honest execution of mixing without disclosing those details. To address this challenge, certain prior studies \cite{valenta2015blindcoin, heilman2017tumblebit} have proposed mixing techniques that aim to shield the mixer entity from knowledge about the involved users. In simpler terms, these mechanisms are structured such that the mixer lacks awareness of the associations between specific inputs and corresponding outputs. \\ 
Blindcoin \cite{valenta2015blindcoin} attained this objective through the utilization of the blind signature scheme outlined by Chaum \cite{valenta2015blindcoin}, effectively concealing linkage details from the mixer.
Heilman et al. \cite{heilman2017tumblebit} introduced a novel approach where both senders and recipients interact with the mixer using cryptographic RSA-based puzzles, ensuring that the mixer entity remains oblivious to the linking information. The high-level schema of TumbleBit is depicted in Figure \ref{fig:TumbleBit}. Further elaboration on these methods is provided in section 3.2.

\paragraph{Off-Chain Transactions.}
Some mixing services employ off-chain transactions to diminish the duration and expenses of the mixing process while enhancing privacy. Due to the inherent untraceability of many off-chain transactions, external observers cannot discern mixing patterns as some transactions are hidden from them.


\paragraph{Fungibility of Inputs.}
Fungibility, within the context of mixing, signifies the requirement that all inputs involved in the mixing process possess equal values. This criterion obfuscates the linkage between senders and receivers in transactions, making the tracing of transactions challenging due to the uniformity of transaction values.

\paragraph{Disconnected Fund Flow.} 
The term "Being Disconnected" denotes severing the complete link between a sender and receiver, preventing any verifiable connection through the transaction graph of the mixer using the maximum-flow algorithm (taint analysis). This can be achieved through swapping methods or by directly delivering cleaned funds to the recipient from an independent address that holds adequate funds. One of the main advantages of maintaining a disconnected mixing graph is the increasing difficulty for a third party to detect mixing transactions since linking the disconnected transactions to each other is a hard task for any entities outside of the mixing process \cite{shojaeinasab2022mixing}.

\paragraph{Address Freshness.}
The freshness of addresses means that every time mixing happens, both the mix's escrow addresses and recipients' output addresses should be fresh addresses created specifically for that mixing process. This is required to prevent connecting the current mixing to other addresses that are not related. Moreover, those mixing services which gives mixing guarantees to the participated users should use fresh and unique addresses in order to create reliable mixing guarantees \cite{bonneau2014mixcoin}. Therefore, both parties should pick addresses with no other possible source of incomes. This helps maintaining the privacy and isolation of the mixing process. It's important to note that this feature is specifically relevant to address-based cryptocurrencies, such as Bitcoin, and isn't applicable within account-based blockchain networks like Ethereum, since the interactions with mixers are based on the mixers' contracts.

\paragraph{Trusted-Execution-Environment.}
A Trusted Execution Environment (TEE) is a secure and separate space within a computer's main working area, which keeps sensitive information and tasks safe from potential threats that could affect the regular parts of the computer. TEEs make sure that only trusted programs can access in the protected space, making it a shielded area for sensitive activities \cite{zhang2016TEEsok}. A TEE could host the mixing process, ensuring that the transaction mixing occurs in a protected and isolated environment with remote attestation (i.e., a third party can verify the correctness of the mixer’s operations, like Intel SGX \cite{anati2013IntelTEE1,mckeen2013IntelTEE2} and ARM TrustZone \cite{arm2009securityTEE}).
In this setup, the TEE securely manages the mixing process, safeguards private keys, and shields transaction details. It also establishes secure communication with the mixer's server, enhancing protection against potential threats. By employing a TEE, users benefit from safer and more private cryptocurrency transactions within the mixer.

\paragraph{Zero-Knowledge proofs.}
Zero-knowledge mixers enable a crucial element of trustless mixing by allowing participants to prove the validity of their transactions without revealing sensitive information, such as the sender, recipient, or transaction amount.
Through ZKPs, users can mathematically confirm that they possess the necessary information to spend their funds, ensuring the integrity of the mixing process while maintaining confidentiality. This approach adds an extra layer of confidentiality and security to the mixing process.

\paragraph{Other techniques.}
Researchers have designed various academic mixing methods primarily leveraging cryptography, distinct from the mentioned functionalities. These methods mostly involve using cryptographic techniques like RSA-based puzzles (in TumbleBit\cite{heilman2017tumblebit} and BSC\cite{heilman2016bsc} papers) and Ring Signatures (in Möbius \cite{meiklejohn2018mobius} paper, notably in Monero cryptocurrency \cite{alonso2020monero}). These methods will be discussed comprehensively in the forthcoming sections.

\section{All the Proposed Mixing Frameworks}\label{sec:allmixings}
Following a comprehensive investigation of the concealment techniques used by many mixing services to ensure funds remain untraceable, this section delves into an examination of the prominent mixing frameworks proposed in the literature. Additionally, we evaluate the practical mixing services that are popular among cryptocurrency users. For our analysis of the mixing frameworks, our focus remains strictly on the foundational papers associated with each, and the discussions in the BitcoinTalk community. On the other hand, when it comes to the widely used practical services, we assess various documented attacks on these platforms and consider the results and analysis found in previous studies.

\subsection{Academic Proposed Mixing Approaches}
In addressing the anonymity concerns within cryptocurrencies, various mixing frameworks have been proposed in academic literature. These efforts, whether decentralized or centralized, have prioritized users anonymity and privacy rather than maximizing profits for the mixer entities. In this section, we delve into the architecture of 8 decentralized and 5 centralized noteworthy mixing frameworks.

\subsubsection{Decentralized Mixing Frameworks}
The core philosophy behind the creation of Bitcoin and other cryptocurrencies was to remove the need for relying on third parties for financial transactions.  Consequently, the cryptocurrency community generally seeks to steer away from centralized services, including mixing services. This has led researchers to suggest decentralized mixing frameworks,  aligning with the fundamental aims of cryptocurrencies, to enhance users' privacy and uphold the primary purpose behind their creation. However, decentralized frameworks generally suffer from a slow mixing process, the need for honest participants majority, and limited scalability.

\paragraph{CoinJoin.}
CoinJoin is a privacy-enhancing technique that allows multiple users to combine their transactions into a single joint transaction \cite{coinjoin}. This process makes it challenging for outside observers to link the original input and output addresses, thus increasing the privacy of participants. CoinJoin is not a complete mixing protocol because it neither describes how participants should be selected, nor how the swapping transaction is actually formed.

\paragraph{CoinShuffle.}
 CoinShuffle was introduced as an enhancement to CoinJoin, presenting it as a fully decentralized system \cite{ruffing2014coinshuffle}. The protocol initiates with the participation of \(n\) users in the mixing procedure. Subsequent to this initiation, each participant generates a new Bitcoin address, intended to serve as their output address in the mixing transaction.

The ensuing step involves the participants shuffling the newly formed output addresses. This shuffling is performed in a manner analogous to a decryption mix network, ensuring that the identities of the address creators remain undisclosed. This procedure unfolds over \(n\) distinct shuffling rounds.

In the initial round, a participant, referred to as Alice, signs her transaction using her private key. She then encrypts this transaction employing the public keys of the remaining \(n - 1\) participants. The resultant encrypted transaction is then relayed to the subsequent participant, Bob. Upon receipt, Bob decrypts the transaction from Alice using his public key. Simultaneously, he signs his personal transaction with his private key, and encrypts it with the public keys of the remaining \(n - 2\) participants. Both transactions (the decrypted one from Alice and his own) are then passed to the third participant.

\begin{figure}
    \centering
    \includegraphics[width=\linewidth]{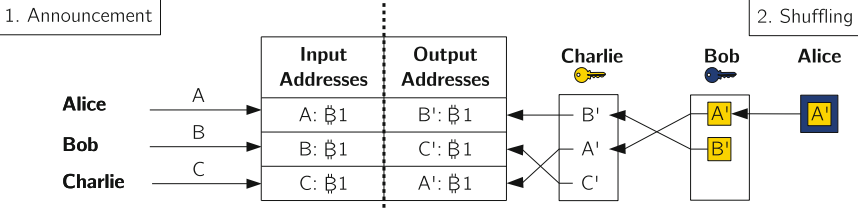} 
    \caption{Overview of CoinShuffle's mixing process \cite{ruffing2014coinshuffle}}
    \label{CoinShuffle}
\end{figure}

This sequence progresses iteratively, culminating when the final participant receives \(n - 1\) encrypted transactions. This last participant decrypts all received transactions using her private key, appends her own transaction to this collection, and then broadcasts the consolidated list. The schema of this protocol is depicted in Figure \ref{CoinShuffle}.

The protocol's success hinges on mutual verification. If participants recognize their respective transactions within the final aggregated list, they will endorse the transaction with their signature. If not, they abstain. However, the main drawback of both CoinJoin and CoinShuffle frameworks is that, since their protocol puts all inputs and outputs in a single transaction, the anonymity set size is limited to the transaction size. Also, the absence of mixing fees makes both Sybil and Dos attacks easy for an attacker (All possible attacks and the resilience of each framework to them are discussed in section \ref{Sec:Attacks} and \ref{sec:Evaluation}).

\paragraph{Xim.}
Bissias et al. put forth Xim as a decentralized P2P mixer that employs a swapping mechanism \cite{bissias2014Xim}. Participants aiming to mix their funds are paired at random and swap their transaction outputs with each other. To mitigate Sybil and DoS attacks, Xim integrates mixing fees. Participation in a mixing action necessitates expenditure from both parties, rendering such attacks prohibitively costly for potential attackers.

\paragraph{CoinParty.}
CoinParty was conceptualized by Ziegeldorf and his colleagues as a mixing strategy that functions through the collaboration of multiple third-party mixing peers \cite{ziegeldorf2015coinparty}. Initially, with \(n\) participating users, \(n\) unique escrow addresses are collaboratively formulated by the mixing peers for users' funds collection. Afterwards, This is followed by the shuffling stage, which is orchestrated similarly to CoinShuffle. Subsequently, the mixing peers sign the shuffled transactions to finalize the mixing process. The jointly generated escrow addresses are the same as the Bitcoin ordinary transactions, resulting in increasing the mixing anonymity. Nonetheless, the absence of mixing fees in this approach diminishes the incentive for mixing peers to work honestly, making the method vulnerable to potential Join-and-Abort attacks.

\paragraph{SecureCoin.}
Ibrahim introduced SecureCoin as a mixing technique that incorporates the Joint-Secret-Sharing protocol \cite{ingemarsson1990JSS} to decrease transaction size while preventing sabotage from mixing peers \cite{ibrahim2017securecoin}. In this design, all senders collaboratively establish a unified escrow address for fund aggregation. Subsequently, output addresses are rearranged in a manner similar to the CoinShuffle protocol. Finally, the senders collectively provide their signatures for the ultimate transaction, enabling the transfer of funds from the aggregation address to the designated outputs.

\paragraph{Möbius.}
Möbius is an Ethereum-based mixing approach within the framework of smart contracts \cite{meiklejohn2018mobius}. This method relies on the Ring-Signature \cite{rivest2001ringSig} scheme and stealth addresses to obfuscate the senders and recipients associations like the strategy observed in Monero \cite{alonso2020monero}. A ring signature is a type of cryptographic digital signature that enables a user to sign a message on behalf of a group (or "ring") without disclosing which individual member's private key was used to create the signature. In terms of mixing, a recipient can withdraw her money from a group of transactions without
revealing the corresponding sender of the withdrawn money.\\
Within the context of Möbius, the process unfolds as follows: Senders furnish both their funds and stealth keys to a designated smart contract that orchestrates the mixing of the provided inputs. Subsequently, each recipient can create a ring signature to withdraw their funds from the contract and transfer them to an ephemeral address, culminating in the completion of the mixing procedure. Notably, the deterministic nature of the Ethereum Virtual Machine (EVM) ensures that the mixing works in a tamper-resistant manner as the entire process occurs on the blockchain, precluding the involvement of any central authority capable of disrupting the integrity of the mixing mechanism.

\paragraph{AMR.}
AMR is a mixer based on zk-SNARKs, aimed at disrupting the traceability connection between coins deposited and withdrawn by a user within a blockchain governed by smart contracts \cite{le2021AMRmixer}. The AMR configuration involves participants depositing a predetermined quantity of coins into a smart contract. Subsequently, this contract establishes a Merkle tree structure over the deposits, generating a commitment that represents the deposited transactions. When a recipient wishes to retrieve their funds from the contract, they are required to validate their familiarity with the committed values associated with specific pre-existing deposit commitments that contribute to the computation of the Merkle tree root. This validation is achieved through the employment of zk-SNARK proofs. Additionally, AMR leverages lending platforms (Like Aave \cite{Aave} and Compound \cite{Compound}) to provide users with the opportunity to accrue interest on their deposited assets. This particular strategy serves as an added motivation for users to retain their funds within the ecosystem.

\paragraph{MixEth.}
Seres et al. introduced MixEth as a decentralized coin mixing solution for Turing-complete blockchains \cite{seres2019mixeth}. The core concept of MixEth revolves around the integration of Neff's verifiable shuffles \cite{neff2001verifiable} within the framework of coin mixing. In the MixEth protocol, participants within the mixer engage in a multi-round shuffling process of their public keys. In each round of shuffling, a shuffler permutes all public keys using Neff's method, then commits them to the mixer along with a Zero-Knowledge proof to prove that the permutation was performed correctly. 
Afterwards, as soon as recipients are confident in the adequacy of the shuffling iterations, they gain the capability to retrieve their assets from the mixing service.

\subsubsection{Centralized Mixing Frameworks}
While decentralization stands as a core objective within the cryptocurrency community, its practical realization is not always optimally efficient. When it comes to mixing, centralized services shows more scalability, speed, and user-friendliness, compared to decentralized ones. Therefore, several researchers have introduced centralized frameworks aimed at enhancing user privacy and anonymity within the services, and mitigate the risk of fund thefts by malicious mixer entities.

\paragraph{MixCoin.}
Bonneau et al. presented MixCoin as the first academic centralized mixer \cite{bonneau2014mixcoin}. Within this framework, a central mixer undertakes the mixing tasks for all participants. When a user deposit their coins into a designated escrow address, the mixer issues a digitally signed message as a guarantee. Should the mixer act maliciously or misappropriate user funds, this guarantee can be publicly disclosed. By leveraging the mixer's public key, the veracity of this guarantee can be ascertained. Consequently, the reputation of a mixer is derived from its track record of honest mixing and user satisfaction. Furthermore, To obfuscate discernible patterns associated with mixing fees, MixCoin adopts a randomized fee strategy. This tactic is especially potent during sequential mixing — when a user sequentially engages multiple mixers. Integrating randomized mixing fees with varied inter-mixing intervals heightens the level of anonymity.

\paragraph{BlindCoin.}
Building upon MixCoin's foundation, Valenta et al. put forth BlindCoin, which leverages chaumian blind signatures \cite{valenta2015blindcoin} and a public ledger to curtail a centralized mixer's insight into the linkage between senders and recipients \cite{chaum1983blind}. This design substantially reduces the susceptibility to permutation leaks. However, it's pivotal to acknowledge that, akin to MixCoin, BlindCoin doesn't offer safeguards against potential fraudulent activities or coin stealing by the mixers.

\begin{figure}
    \centering
    \includegraphics[width=0.75\linewidth]{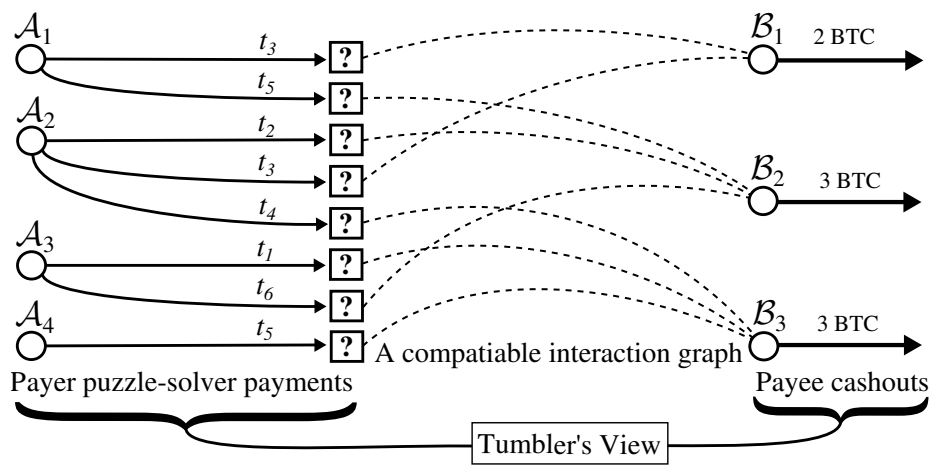}
    \caption{Overview of TumbleBit's mixing graph \cite{heilman2017tumblebit}. Tumbler's view on mixing associations is blinded.}
    \label{fig:TumbleBit}
\end{figure}

\paragraph{TumbleBit.}
TumbleBit was proposed as a centralized mixing approach aimed at preventing connections between senders and receivers \cite{heilman2017tumblebit}. In this method, the sender (Alice) and the receiver (Bob) engage with a central entity called a tumbler $\tau$. First, Alice and Bob establish payment channels with $\tau$. Then, Alice creates a 2-of-2 escrow transaction denoted as
$T_{escr}$($A$, $\tau$), allowing $\alpha$ bitcoins to be accessed through a joint signature from both Alice and $\tau$. Similarly, the tumbler forms another shared transaction denoted as $T_{escr}$($\tau$, $B$), and these transactions should be committed to the blockchain. Afterwards, Alice and Bob interact with the tumbler off-chain, employing RSA-based puzzles. These interactions facilitate Bob for receiving a signature from $\tau$ for his transaction $T_{escr}$($\tau$, B), and help $\tau$ receiving Alice's signature of her transaction $T_{escr}$($A$, $\tau$). Finally, $\tau$ make a transaction from $T_{escr}$($A$, $\tau$) to his fresh generated address, and Bob makes another transaction from $T_{escr}$($\tau$, $B$) to his own fresh generated address, and the mixing process finished. Additionally, TumbleBit employs a chaumian blind signature \cite{chaum1983blind} scheme to safeguard the tumbler from learning about participant associations, as it shown in Figure \ref{fig:TumbleBit}.

\paragraph{BSC.}
Heilman and colleagues introduced Blindly Signed Contracts (BSC) as a mixing approach that capitalizes on \emph{blind signatures} principles akin to those in TumbleBit. This method is designed to establish unlinkability in both \emph{on-chain} and \emph{off-chain} contexts. In the on-chain process, the confirmation of all mixing transactions on the blockchain results in a comparatively slow mixing speed. On the other hand, the \emph{off-chain} strategy leverages micro-payment channel networks \cite{poon2016lightningNetwork, decker2015paymentChannel} to keep some mixing transactions off the primary blockchain. This is advantageous as \emph{off-chain} transactions exhibit enhanced speed, scalability, and the potential for reduced fees compared to the conventional \emph{on-chain} transactions.

\paragraph{Obscuro.}
Obscuro is a mixing protocol that capitalizes on the capabilities of a Trusted Execution Environment (TEE), elaborated upon in section 3.1. This paradigm allows participants to verify the isolated functionalities through remote attestation, assuring them of a robust mixing set size \cite{tran2018obscuro}. The protocol's efficacy was evaluated within the Bitcoin Testnet by its authors, and its adaptability was demonstrated through a smart contract implementation on the Ethereum network \cite{ObscuroWhitePaper}.

\begin{table}[htbp]
    \centering
    \caption{Functionalities employed by all mixing solutions proposed in academia.\tablefootnote{In the following cells of all tables, "a." stands for "applicable", "n.a." stands for "not applicable", and "-" means that these is not adequate information to determine the corresponding value.}}
    \setlength\tabcolsep{5pt}
    \begin{tabular}{l|ccccccccccccccc}
        \toprule
        & \rotatebox{80}{Swapping} & \rotatebox{80}{Aggregating} & \rotatebox{80}{Peeling Chain} & \rotatebox{80}{Splitting} & \rotatebox{80}{Chain Hopping} & \rotatebox{80}{Random Fee} & \rotatebox{80}{Random Delay} & \rotatebox{80}{Third-Party Blinding} & \rotatebox{80}{Off-Chain} & \rotatebox{80}{Input Fungibility} & \rotatebox{80}{Disconnected Fund Flow} & \rotatebox{80}{Address Freshness} & \rotatebox{80}{TEE} & \rotatebox{80}{ZKP} & \rotatebox{80}{Others} \\
        \midrule
        CoinJoin        & \cmark & \xmark & \xmark & \xmark & \xmark & no fee & a. & n.a. & \xmark & a. & \xmark & a. & \xmark & \xmark & \xmark \\
        CoinShuffle     & \cmark & \xmark & \xmark & \xmark & \xmark & no fee & a. & n.a. & \xmark & a. & \xmark & \cmark & \xmark & \xmark & \xmark \\
        Xim             & \cmark & \xmark & \xmark & \xmark & \xmark & \xmark & a. & n.a. & \xmark & a. & \cmark & a. & \xmark & \xmark & \xmark \\
        CoinParty       & \cmark & \xmark & \xmark & \xmark & \xmark & no fee & a. & n.a. & \xmark & a. & \xmark & a. & \xmark & \xmark & \xmark \\
        SecureCoin      & \cmark & \cmark & \xmark & \xmark & \xmark & a. & a. & n.a. & \xmark &  a. & \xmark & a. & \xmark & \xmark & \xmark \\
        Möbius          & \xmark & \xmark & \xmark & \xmark & \xmark & a. & a.\tablefootnote{\label{applicableDelayNote1}Since withdrawing money from Ethereum-based mixing services can be done whenever users choose to (due to the account-based nature of the Ethereum network), users can add random delays by their own.} & n.a. & \cmark & \cmark & \xmark & n.a. & \xmark & \xmark & \cmark \\
        AMR             & \xmark & \xmark & \xmark & \xmark & \xmark & a. & a.\footref{applicableDelayNote1} & n.a. & \xmark & \cmark & \xmark & n.a. & \xmark & \cmark & \xmark \\
        MixEth          & \cmark & \xmark & \xmark & \xmark & \xmark & a. & a.\footref{applicableDelayNote1} & n.a. & \cmark & \cmark & \xmark & n.a. & \xmark & \cmark & \xmark \\
        Mixcoin         & a. & a. & a. & a. & \xmark & \cmark & a. & \xmark & \xmark & a. & a. & \cmark & \xmark & \xmark & \xmark \\
        BlindCoin       & a. & a. & a. & a. & \xmark & \cmark & a. & \cmark & \xmark & a. & a. & \cmark & \xmark & \xmark & \xmark \\
        TumbleBit       & \xmark & \xmark & \xmark & \xmark & \xmark & a. & a. & \cmark & \cmark & a. & \cmark & a. & \xmark & \xmark & \cmark \\
        BSC             & \xmark & \xmark & \xmark & \xmark & \xmark & a. & a. & \cmark & \cmark & a. & \cmark & a. & \xmark & \xmark & \cmark \\
        Obscuro         & a. & a. & a. & a. & \xmark & a. & a. & \xmark & \xmark & a. & a. & a. & \cmark & \xmark & \xmark \\
        \bottomrule
    \end{tabular}
    \vspace{4pt}
    \label{tab:AcademiaMixers}
\end{table}

Table \ref{tab:AcademiaMixers} outlines the functionalities utilized by academic mixers to obfuscate transactions and facilitate the mixing process. Notably, several academic solutions incorporates swapping methods by permuting the associations of inputs and outputs, in contrast with aggregating and peeling chain, which weren't signified in academic approaches. On the other hand, certain methodologies such as Möbius, TumbleBit, and BSC utilize cryptographic puzzles such as ring signatures and RSA-based puzzles. Additionally, Obscuro showcased an innovative application of a Trusted Execution Environment, establishing itself as a pioneering approach within this domain. AMR and MixEth also leveraged ZKPs in samrt-contract-based cryptocurrencies. It's important to note that the majority of additional obfuscation techniques, such as randomizing fees/delays and employing fresh addresses, can be implemented within these mixing frameworks, even though they were not explicitly mentioned in the papers. Notably, academic frameworks have not extensively addressed Cross-Chain mixing, despite its significance in the evolving world of decentralized finance.

\subsection{Real-World Mixing Approaches}
In this section, we will conduct a comprehensive examination of eighteen commonly employed mixing services accessible within the market. While certain services discussed below may have become outdated by the time of this paper's publication, they merit analysis due to the significant volume of mixing-related data associated with them.\\

Regarding centralization and decentralization in practical services, a significant portion of them lean towards centralization, especially within the Bitcoin ecosystem. However, this term is completely different in the realm of smart-contract-based cryptocurrencies like Ethereum, Owing to their capacity to execute Turing-Complete algorithms. To maintain focus on the core subject of this section, discussions related to this aspect have been moved to Section \ref{comparison}. Within this section, the first sixteen services are centralized, and the last two services (Join-Market and Tornado Cash) are decentralized. Given that other decentralized services either lack the necessary robustness or lack adequate reputations, we only discuss those two remarkable services. To the date of this publication, we have not found any reputable service that has more powerful functionalities compared to the mentioned services.

\paragraph{BitcoinFog.}
Bitcoin Fog stands as one of the earliest centralized bitcoin mixing services, which was exclusively accessible through the Tor network. This service permits the creation of a maximum of five addresses for depositing bitcoins and takes a (random) fee between 1–3\% of the transaction value. The withdrawal process allows for dispersing bitcoins to a maximum of twenty addresses, spanning a timeframe of 6 to 96 hours. In a study conducted by Moser \cite{moser2013mixer1}, an evaluation of this mixer was undertaken by initiating transactions through it and subsequently analysing the resulting transaction graph created by the mixing service. \\
The graph generated from this evaluation is visualized in Figure 3. Bitcoin Fog aggregates incoming transactions into distinct aggregation addresses referred to as "communities". Notably, some of these communities contain as much as 50,000 BTC. Following this aggregation, output transactions are derived from these communities and directed toward their designated recipients. Noteworthy is Bitcoin Fog's utilization of randomized fees and varying mixing delays, strategically implemented to hinder external attempts at detecting the mixing graph by analyzing fee structures and transaction timing patterns.\\
It's important to note that BitcoinFog does not mandate participants to send a predetermined sum of money to the mixing service. Consequently, if a user transfers funds that significantly deviate from the average of incoming funds to the mixer, there is a heightened likelihood of detection due to the small anonymity set provided. \\
Notably, the operator of this service was arrested in 2021 for allegedly running this service, which was considered the longest-running bitcoin money laundering service on the darknet \cite{BitcoinFogSeizureReport}.

 \begin{figure}
    \centering
    \includegraphics[scale=0.6]{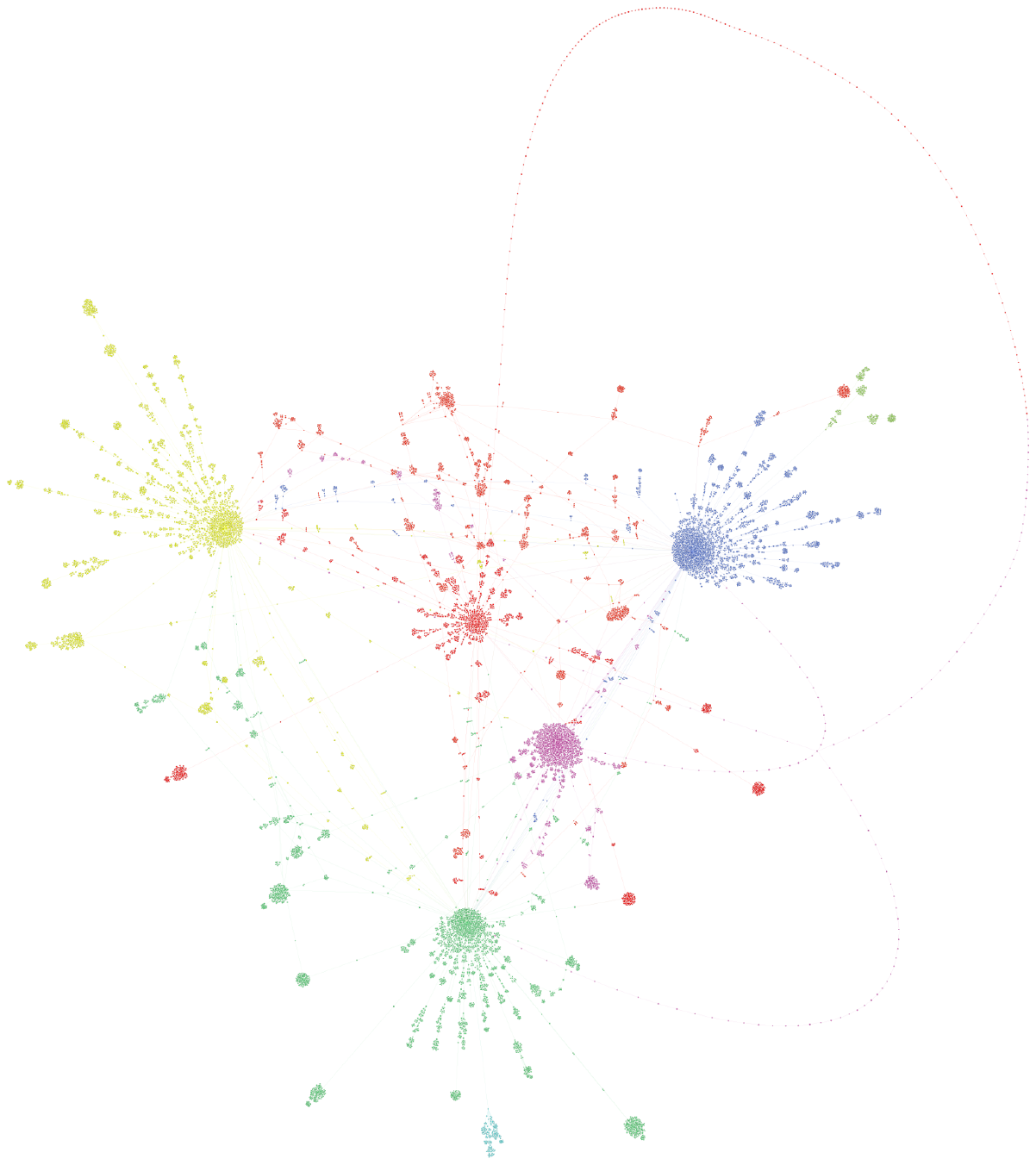}
    \caption{Sample of created communities (aggregation addresses) in BitcoinFog \cite{moser2013mixer1}}
    \label{fig:enter-label}
\end{figure}

\paragraph{BitLaundry.}
BitLaundry represents an uncomplicated mixing service that differs from Bitcoin Fog in its operational methodology. Unlike Bitcoin Fog, BitLaundry doesn't facilitate the deposition of bitcoins into a virtual wallet. Instead, the service necessitates the specification of destination addresses, the count of outgoing transactions, and a designated time span. This prompts the generation of a one-time-use address, to which a minimum threshold of coins must be sent by the user. The fee structure of BitLaundry's mixing service is split into to part. The first part constitutes 2.49\% of the total amount, while the second part involves a charge of a static amount for each outgoing transaction.\\
Based on Moser's evaluation in 2013 \cite{moser2013mixer1}, it was observed that BitLaundry employed fund aggregation as a technique to obscure associations. However, Moser's analysis indicated that due to the limited number of participants within the mixing service, certain inputs were directly linked to corresponding outputs. Consequently, the mixing process exhibited a reduced level of anonymity.

\paragraph{Blockchain.com.}
Blockchain.com, a renowned cryptocurrency-focused website and platform, predominantly offers an array of services pertaining to digital currencies, with a primary emphasis on Bitcoin. Notably, within the Blockchain.com wallet's historical features, there existed a service known as "Send Shared". This service operates by facilitating Bitcoin exchanges among various users through a shared wallet structure. During its operation in 2013, it levied a modest mixing fee of 0.5\%. At that time, it stood out as an economical option in the landscape of cryptocurrency services. \\
Moser's examination \cite{moser2013mixer1} revealed that this service employed an aggregation technique to obscure the mixing process. Notably, some of the aggregating addresses contained substantial sums, approximately 2,000 BTC each. In their examination, they made eleven transactions to this service and detected eight individual clusters of aggregating addresses, which shows that the mixer is resilient to the taint analysis attack. Following the generation of these aggregating addresses, funds underwent a process of division into several smaller transactions, and these were subsequently subdivided until they reached their designated destination addresses.

\paragraph{DarkLaunder, Bitlaunder and CoinMixer}
According to the evaluation conducted by Balthasar et al., in their 2017 study, these particular services were identified as the least effective mixing services at the time of assessment\cite{balthasar2017MixerAnalysis2}. The authors chose to examine them together due to their belief that these services were likely under common ownership and displayed striking similarities in their operations and attributes. Consequently, an analysis of these services can offer valuable insights into the pitfalls that mixing services should steer clear of in order to attain a higher level of anonymity. \\
First problem of these mixers comes from that they require users to create account for having access to the mixer. Furthermore, they stored this user information alongside the IP addresses associated with transaction requests, as well as comprehensive transaction details in their database. What exacerbates this concern is the vulnerability of their websites, making it relatively simple for potential attackers to gain access to this stored data. \\
Moving on to their obfuscating techniques, they employed an aggregation approach akin to several other services. However, a significant drawback was the aggregation addresses were static for a long time, which led to a concentration of the mixing process and made it discernible to potential attackers. Additionally, these services encountered a scarcity of participants, resulting in notably small anonymity set size and poor obfuscating permutations. In summary, these services exhibited exceedingly low levels of security and privacy, rendering them illustrative examples of common pitfalls to be noticed when implementing a secure mixer.

\paragraph{Helix.}
Helix, a cryptocurrency tumbler, gained prominence within the dark web community due to its affiliation with the creators of the renowned dark web search engine, Grams. The developers assert that Helix utilizes innovative proprietary technology to not only cleanse Bitcoins but also generate entirely new ones untouched by the darknet's transactions. This service was introduced to the public in June 2014. It is worth noting, however, that Helix's founder later pleaded guilty to a conspiracy charge involving the laundering of approximately \$300 million. The potential consequences for this individual may include a twenty-year prison sentence and a fine of either \$500,000 or twice the value of the assets involved in the illicit transaction, as indicated in the references \cite{HelixGuilty1, HelixGuilty2}.\\
Helix is accessible only using Tor, and it utilizes random time delays for mixing to achieve anonymity. It takes 2.5\% of the transaction value as a mixing fee, and only allows withdrawals of 0.02 BTC or more\footnote{\label{footNote2}This data was captured in 2016}. Based on the assessment of Balthasar et al. \cite{balthasar2017MixerAnalysis2}, Helix also uses an aggregating method to accumulate funds into an address, and then distribute them to the recipients using a peeling chain. However, their observation shows that wallets and withdrawals of multiple customers are present on the same transaction, or very close to each other, thus it is easy to identify them. In short, although linking senders to recipients is a hard challenge in Helix, it is easy to verify that their transactions were passed from a mixer, by performing a taint analysis.

\paragraph{Alphabay.}
AlphaBay was a darknet market operating from the year 2014, a year after the Silk Road market was busted in 2013. In 2017, this platform was forcefully terminated, and its principal administrator was arrested \cite{AlhphaBayCrimeReport} . However, this service was relaunched in August 2021 by the self-described co-founder of the AlphaBay. Prior to its 2017 shutdown, AlphaBay's management utilized a mixer to obscure the origins of the market's financial transactions.
Since Alphabay had approximately 400,000 users in 2017, a significant proportion of users employed the mixer for their monetary transactions. Consequently, the mixer had a large mixing anonymity set, making it an interesting subject for studying with the potential to yield valuable insights about its operation mechanism.
Balthasar et al. \cite{balthasar2017MixerAnalysis2} assessed this mixing service by creating 35 different transactions to this mixer, and found three main clusters of aggregated funds, which shows that although this mixer had a multitude of users, it did not use a large number of clusters and its mixing transactions can be found easily by performing a taint analysis. This mixer also used other third-party mixing services like Helix, in order to make the mixing process more anonymous. Also, it utilizes random delays between creating mixing transactions to prevent any time-based mixing detection.

\paragraph{MixTum}
MixTum \cite{MixTumWebsite}, established in 2018, claims to have a separate pool of Bitcoin from cryptocurrency stock exchanges such as Binance, OKEex, and DigiFinex, which is used to obfuscate users' funds with external coins from exchanges. Shojaeinasab et al. \cite{shojaeinasab2022mixing} have assessed the mixing mechanism of MixTum, revealing its similarity to typical mixers in aggregating incoming funds into an aggregation address and subsequently distributing them through a peeling chain. Furthermore, this platform ensures privacy enhancement by avoiding the aggregation of transactions from a single address into a common aggregation address, and also leveraging fee and delay randomization. \\
Additionally, MixTum furnishes users with a PGP-signed guarantee letter containing details about the mixing process, and its PGP fingerprint is publicly accessible on its website. This provision enables users to validate any concerns or complaints regarding improper mixing of funds from the service. This method is similar to MixCoin \cite{bonneau2014mixcoin} guarantee letter, with the distinction that MixCoin signs its guarantee letter using its Bitcoin private key instead of a PGP fingerprint.

\paragraph{CryptoMixer.}
In 2016, CryptoMixer \cite{CryptoMixerWebsite} was introduced as a Tor-accessible Bitcoin mixing service. This service functions by aggregating user funds into aggregation addresses, followed by their distribution to recipients using peeling chains \cite{shojaeinasab2022mixing}. Additionally, CryptoMixer offers a guarantee letter, akin to the concept introduced in the MixCoin paper \cite{bonneau2014mixcoin}, signed using the mixer's private key. This feature allows users to publish their complaints from the mixer using a valid evidence. Moreover, CryptoMixer provides customizable obfuscation options, including the number of input and output addresses, mixing delay, and service fee. Furthermore, upon sending funds to the mixer, users are assigned a unique identification code. This code ensures that a user's coins are not mixed with those from previous transactions in subsequent dealings. It is worth noting, however, that there is a lack of documented information regarding CryptoMixer's operational details \cite{balthasar2017MixerAnalysis2}.

\paragraph{Blender.}
In 2017, Blender\cite{BlenderWebsite} was introduced as a centralized mixing service, and it employed several techniques, such as tailored mixing fees, mixing delays, and distinctive mixing codes, much like CryptoMixer, to establish a level of anonymity. Furthermore, Blender required that the input address and at least one of its output addresses be of the P2SH \footnote{P2SH: Pay-to-Script-Hash} type. This mixer's mixing process is claimed to resemble MixTum and CryptoMixer, involving the consolidation of input funds into aggregation addresses and the utilization of peeling chains to distribute mixed funds to recipients \cite{shojaeinasab2022mixing}.

\paragraph{ChipMixer.}
Established in 2017, ChipMixer quickly gained popularity as a top-notch mixing service. However, in March 2023, the FBI seized the operation for processing over \$3 billion in illegal transactions \cite{ChipMixerSeizureReport}.
ChipMixer operates by utilizing units known as "chips." Users deposit their funds into the mixer and receive "chips" of a pre-specified value, consisting of bitcoins from various people. The “chip” is basically just BTC in a new address for which the participants are provided the private key, and chips are always created before a user deposits their funds into the mixer. Interestingly, Mixing fees are completely donation-based in this system, with a "Pay What You Want" strategy, contributing to its popularity as a free mixing service. Between 2017 and 2020, ChipMixer is estimated to have laundered about 53,000 bitcoins \cite{pakki2021everything}.

As the chips are fungible, tracing the funds in this system is a challenging task. Researchers like Pakki et al. \cite{pakki2021everything} and Wu et al. \cite{wu2021towards} examined this service by making several transactions through it. Their findings suggest that the service pools funds into a few aggregation addresses divides the funds into many interchangeable addresses (chips), and then employs multi-path mixing to distribute these chips to users.

\paragraph{Wasabi Wallet.}
Wasabi Wallet \cite{WassabiWalletWebsite} is a publicly available Bitcoin wallet designed for desktop computers, emphasizing user privacy and security. It incorporates a trustless mixing technique, akin to the CoinJoin method. When users initiate transactions within this wallet, Wasabi executes a CoinJoin transaction, generating distinct groups of funds with varying denominations. Subsequently, funds from these generated groups are transferred to the intended recipients \cite{wu2021towards}. The transaction cost for utilizing Wasabi consists of the standard mining fee plus an additional 0.3\% fee.

\begin{figure} 
  \centering
  \begin{minipage}{0.45\textwidth}
    \centering
    \includegraphics[width=\linewidth]{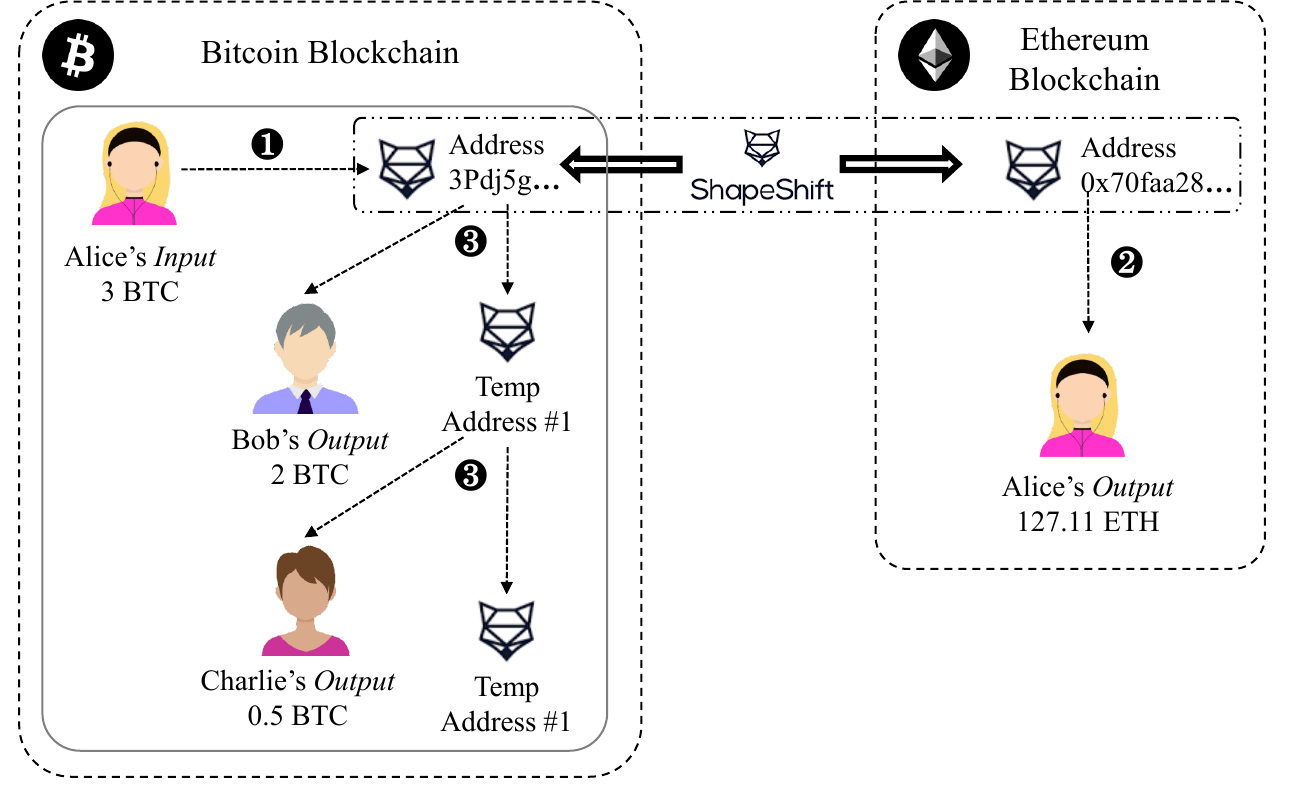}
    \caption{Overview of ShapeShift's mixing process \cite{wu2021towards}}
    \label{fig:ShapeShift}
  \end{minipage}
  \hfill 
  \begin{minipage}{0.45\textwidth}
    \centering
    \includegraphics[width=\linewidth]{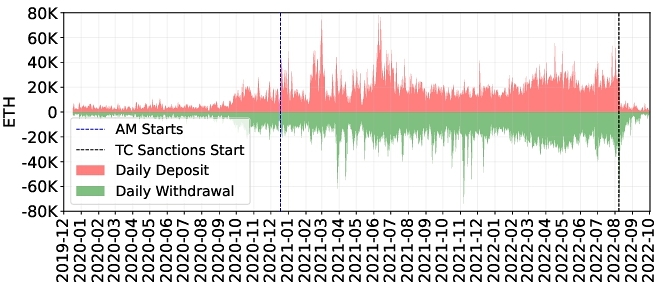}
    \caption{Daily transactions in TC ETH pools. There was a
panic exit when the OFAC sanctions were announced \cite{wang2023zkpMixersEval}}
    \label{fig:TCstat}
  \end{minipage}
\end{figure}

\paragraph{ShapeShift.}
ShapeShift \cite{ShapeShiftWebsite}, founded by Erik Voorhees in 2014, is a cryptocurrency exchange platform that emphasizes privacy. It allows users to trade different cryptocurrencies without revealing their identities, making it suitable for exchanging various cryptocurrencies across different blockchain networks. However, although this service was not designed for a mixing purpose, it can be used for cross-chain mixing due to its provided coin exchanging anonymity. To illustrate, as reported by the Wall Street Journal, this service has been linked to facilitating at least \$9 million in money laundering activities over several years
\cite{ShapeShift+WallStreet+Report}.
Consequently, in response to mounting pressure from governments and media outlets, ShapeShift applied the Know-Your-Customer (KYC) policy and it mandates account creation for accessing its services. Also, ShapeShift stated that they utilize anti-money laundering methods in their system, and rejected the mentioned accused report.\\
Figure \ref{fig:ShapeShift} illustrates an example of using this service as a mixer. First, Alice sends 3 BTC to ShapeShift and receives 127.11 Ether in Ethereum later. Since these chains are totally different from each other, there is not any direct link between senders and receivers, and Alice can use the exchanged Ethers as a mixed fund. Afterwards, the Bitcoin sent by Alice will be organized as a peeling chain to distribute Bitcoins to other users (e.g., Bob and Charlie in this figure) who swap other cryptocurrencies for Bitcoin \cite{wu2021towards}.


\paragraph{Bitmix.}
Bitmix.bz \cite{Bitmix.bzWebsite} was introduced in 2017, and currently serves mixing for Bitcoin, Dash, and Litecoin. This service claimed to utilize several methods like dust-attack prevention, randomized fees and delays, and guarantee letters in order to achieve anonymity. The mixing mechanism of this service is almost like most of the previous described services; aggregating funds into some addresses, then using a peeling chain for funds distribution \cite{wu2021towards}.  

\paragraph{Sudoku Wallet.}
Sudoku Wallet was announced in 2019 within the bitcointalk forum. The service is a single-use wallet that outputs private keys rather than on-blockchain transactions. As the creators described, the mixing fees are randomized from 0.5\% to 1\% plus the CoinJoin fee and there is not any database and no log is saved in their system. Furthermore, they asserted that this service performs CoinJoin transactions using its incoming funds and sends them to newly generated addresses. Afterwards, the private keys of the corresponding addresses are given to the users to spend them \cite{SodukuWalletPubPage}. \\
Although the creators of this service explained their mixing protocol, the interaction of Pakki et al. \cite{pakki2021everything} with this mixer showed different results from the creators' claim. After transacting 3 times with the mixer and analyzing transactions using the Chainalysis tool, they did not identify any evidence of CoinJoin transactions. Also, the mixing fees of their transactions were inconsistent, as one of them had a mixer fee of 0 BTC while another had a fee of 90\%. Consequently, it can be inferred that this mixer does not offer enough safety and reliability, besides its poor implementation.  

\paragraph{JoinMarket.}
JoinMarket \cite{JoinMarketWebPage} is a decentralized peer-to-peer (P2P) mixing service designed to connect users seeking to engage in CoinJoin transactions. This platform consisted of two primary participant categories: Market Makers and Market Takers. Market Makers are responsible for crafting CoinJoin transactions, while Market Takers are individuals seeking to mix their funds. Takers pay a fee to makers which works as an incentive, and notably, all actions within the system occur without reliance on any centralized authority. JoinMarket is remarkable within the Bitcoin network as one of the scarce decentralized mixing services, widely adopted by users seeking to obfuscate the origins of their funds.

\paragraph{Tornado Cash.}
Tornado Cash is an open-source software project on the Ethereum Network launched in December 2019 \cite{pertsev2019tornadoCash}, presently stands as the predominant mixer within the Ethereum ecosystem. When a user wishes to blend their assets, they transfer their tokens into the Tornado Cash smart contract. The funds are secured using a Merkle tree cryptographic commitment scheme, hiding both the amount and source of the funds. Later, when taking funds out of the Tornado pool, a withdrawer proves the ownership of a commitment without revealing which specific commitment is being spent by using zero-knowledge proofs. This provides unlinkability between the deposited and withdrawn funds.\\
In August 2022, the United States Department of the Treasury's Office of Foreign Assets Control (OFAC) imposed sanctions on Tornado Cash due to its involvement in laundering a sum exceeding \$7 billion \cite{TornadoCashSanctionLink}. After that, OFAC added the Tornado Cash website and addresses used this mixer to the "Specially Designated Nationals And Blocked Persons" list, thereby prohibiting U.S. citizens from engaging in any transactions involving the Tornado Cash website or addresses listed in the sanctions. Afterwards, the rate of using Tornado Cash dropped significantly, as depicted in Figure \ref{fig:TCstat}. It is noteworthy that a subset of Tornado Cash users have tried to diffuse banning from OFAC to other addresses by performing a dusting attack. Wang et al. conducted an assessment of the OFAC's sanctions' effects on ZKP-based mixers \cite{wang2023zkpMixersEval}. \\ \\

\begin{table}[htbp]
    \centering
    \caption{Functionalities employed by all mixing services available in market}
    \setlength\tabcolsep{5pt}
    \begin{tabular}{l|ccccccccccccccccc}
        \toprule
         & \rotatebox{85}{Birth Date} & \rotatebox{85}{Active?} & \rotatebox{85}{Centralized} & \rotatebox{85}{Swapping} & \rotatebox{85}{Aggregating} & \rotatebox{85}{Peeling Chain} & \rotatebox{85}{Splitting} & \rotatebox{85}{Chain Hopping} & \rotatebox{85}{Random Fee} & \rotatebox{85}{Random Delay} & \rotatebox{85}{3rd-Party Blinding} & \rotatebox{85}{Off-Chain} & \rotatebox{85}{Input Fungibility} & \rotatebox{85}{Disconnected Fund Flow} & \rotatebox{85}{Address Freshness} & \rotatebox{85}{TEE} & \rotatebox{85}{ZKP}
        \\
        \midrule
         BitcoinFog                     & 2013 & \xmark & \cmark & \xmark & \cmark & \cmark &  -  & \xmark & \cmark & -      & \xmark & - & \xmark & - & \cmark & \xmark & \xmark
         \\
         BitLaundry                     & 2013 & \xmark & \cmark & \xmark & \cmark & -      &  -   & \xmark & \xmark & \xmark & \xmark & -& \xmark & - & \cmark & \xmark & \xmark 
         \\                             
         Blockchain.com                 & 2013 & \xmark & \cmark & \xmark & \cmark & -      & \cmark & \xmark & \xmark & -      & \xmark & - & \xmark & \cmark & -  & \xmark & \xmark
         \\
         DarkLaunder                    & 2015 & \xmark & \cmark & \xmark & \cmark & \cmark & \xmark & \xmark & \xmark & \xmark & \xmark & - & \xmark & \xmark & \xmark & \xmark & \xmark 
         \\
         Bitlaunder                     & 2015 & \xmark & \cmark & \xmark & \cmark & \cmark & \xmark & \xmark & \xmark & \xmark & \xmark & - & \xmark & \xmark & \xmark & \xmark & \xmark
         \\
         CoinMixer                      & -    &   -    & \cmark & \xmark & \cmark & \cmark & \xmark & \xmark & \xmark & \xmark & \xmark & - & \xmark & \xmark & \xmark & \xmark & \xmark
         \\
         Helix                          & 2017 & \xmark & \cmark & -      & \cmark & \cmark & -      & \xmark & \xmark & \cmark & \xmark & - & \xmark & \xmark &  -     & \xmark & \xmark
         \\
         Alphabay                       & 2017 & \cmark & \cmark & \xmark & \cmark & \cmark & -      & \xmark & \xmark & \cmark & \xmark & - & \xmark & \xmark &  -     & \xmark & \xmark
         \\
         MixTum                         & 2018 & \cmark & \cmark & \xmark & \cmark & \cmark & \xmark & \xmark & \cmark & \cmark & \xmark & - & \xmark & \xmark & \cmark & \xmark & \xmark
         \\
         CryptoMixer                    & 2016 & \cmark & \cmark & \xmark & \cmark & \cmark & -      & \xmark & \cmark & \cmark & \xmark & - & \xmark & \xmark & \cmark & \xmark & \xmark
         \\
         Blender                        & 2017 & \cmark & \cmark & \xmark & \cmark & \cmark & -      & \xmark & \cmark & \cmark & \xmark & - & \xmark & \xmark & \cmark & \xmark & \xmark
         \\
         ChipMixer                      & 2017 & \xmark & \cmark & \xmark & \cmark & -      & \cmark & \xmark & a.     & \cmark & \xmark & - & \xmark & \xmark & \cmark & \xmark & \xmark 
        \\
         Wasabi Wallet                 & 2018 & \cmark & \cmark & \cmark & \xmark & \xmark & \cmark & \xmark & \xmark & -      & \xmark & - & \xmark & \xmark &  -      & \xmark & \xmark
         \\
         ShapeShift                     & 2014 & \cmark & \cmark & \cmark & \xmark & \cmark & \xmark &
                                        \cmark & -      & -      & \xmark & - & \xmark & \cmark &
                                        \cmark & \xmark & \xmark
         \\
         Bitmix.bz                      & 2017 & \cmark & \cmark & \cmark & \cmark & \cmark & \cmark & \xmark & \cmark & \cmark & \xmark & - & \xmark & \xmark & \cmark & \xmark & \xmark
         \\
         Sudoku Wallet                  & 2019 & \xmark & \cmark &   -    & -      & -      & -      & \xmark & \cmark & \cmark & \xmark & - & \xmark & -    & -      & \xmark & \xmark
         \\
         Join-Market                    & 2017 & \cmark & \xmark & \cmark & \xmark & \xmark & \xmark & \xmark & a. & a.\tablefootnote{\label{applicableDelayNote}Since performing a CoinJoin transaction in Join-market and withdrawing money from Tornado Cash can be done whenever users choose to, users can add random delays by their own.} & n.a. & \xmark & \xmark & \xmark & a. & \xmark & \xmark  
         \\
         Tornado Cash                   & 2019 & \cmark & \xmark & \xmark & \xmark & \xmark & \xmark & \xmark & \xmark & a.\footref{applicableDelayNote} & n.a. & \xmark & \xmark & \xmark & n.a. & \xmark & \cmark \\
         
    \bottomrule
    \end{tabular}
    \vspace{4pt}
    \label{tab:MarketMixers}
\end{table}

Table \ref{tab:MarketMixers} outlined functionalities utilized by real-world mixing services, and as depicted, most services are centralized. A repetitive pattern in most services involves using aggregation techniques to accumulate funds into a designated address and using a peeling chain to distribute these funds to individual users. This approach is favored over alternatives such as swapping, diverging from what is typically observed in academic frameworks. Also, third-party blinding has not been used by services, and it is clear that mixer entities do not want to blind themselves from the mixing information of their users. Moreover, Since forcing users to deposit coins only in a specific denomination to achieve fungibility is not desirable for most users, this functionality is not used by any service. However, a solution for users to heighten the obfuscation for their mixed funds is to fragment their output transactions to align with the average funds sent to the mixer by all users \cite{moser2013mixer1}. In addition, it is worth noting that several services may use off-chain transactions in platforms like payment channel networks, but it is a challenging task for an external observer to ascertain the utilization of off-chain transactions, as they occur in a separate layer distinct from the main Bitcoin network transactions. 

\subsection{Comparison Between Academic and Real-World Services}\label{comparison}
By examining the data in Table~\ref{tab:AcademiaMixers} and Table~\ref{tab:MarketMixers}, it is clear that academia and market display considerable differences, which are mostly attributable to their different goals for the creation of mixing services. 

In the context of Bitcoin mixing services, since the Bitcoin scripting language is not Turing-Complete, it is necessary to construct mixers in a higher layer of the Bitcoin network. Notably, most academic endeavors have focused on developing decentralized mixing approaches to achieve the fundamental objective of cryptocurrencies, which is to minimize regulatory control exerted by centralized entities. Various techniques such as multi-round digital signing (e.g., CoinJoin \cite{coinjoin}, CoinShuffle \cite{ruffing2014coinshuffle}) and Joint-Address-Creation (e.g., CoinParty \cite{ziegeldorf2015coinparty}, SecureCoin \cite{ibrahim2017securecoin}) have been applied within decentralized mixing frameworks to achieve this objective.

Even in centralized approaches, there have been novel innovations for blinding the centralized mixer party from having any information about associations between inputs to outputs, by leveraging cryptographic schemes like Blind Signature which were used in TumbleBit \cite{heilman2017tumblebit} and BSC \cite{heilman2016bsc}. Moreover, Obscuro \cite{tran2018obscuro} suggested utilizing Trusted-Execution-Environment (explained in section \ref{sec:3-functionalities}) to monitor the mixing workflow which is performed by a mixer entity. 

However, in practical mixing services found in real-world applications, most of the methods discussed in academic papers have not been fully implemented. As outlined earlier, these services generally aggregate funds into an address and then distribute them using a peeling chain \cite{shojaeinasab2022mixing, moser2013mixer1, balthasar2017MixerAnalysis2, wu2021towards}. This approach is commonly used perhaps because it is a well-established and straightforward technique to implement. Furthermore, there hasn't been any significant attack on this method that can link inputs and outputs to each other. Additionally, a mixer can effortlessly add its own supply funds to the aggregation addresses, enhancing the anonymity pool. These steps are easy to develop by the services, and there's no need for complex implementations, resulting in a widespread and common method to use in mixing services. Also, centralized services tend to avoid blinding themselves from the mixing data, and we haven't come across any service employing blinding techniques to obscure sender-recipient relationships.

Nevertheless, despite this gap, several services have adopted academic methods to assure their users of their reliability and enhance the security of their mixing process. Specifically, the MixCoin guarantee letter \cite{bonneau2014mixcoin} has been incorporated into multiple mixers, and some services have integrated features like fee and delay randomization to prevent attackers from identifying mixing transactions based on fee or time patterns.

To talk about other services that use alternative methods (rather than aggregating), some services like Wasabi Wallet \cite{WassabiWalletWebsite} act as an intermediary responsible for collecting mixing transactions from users and executing a CoinJoin transaction (which is decentralized in its nature) in a centralized manner. Conversely, Join-Market \cite{JoinMarketWebPage} introduced a decentralized P2P network, enabling participants to generate CoinJoin transactions in a decentralized way. Additionally, leveraging anonymous cryptocurrency exchanges for cross-chain mixing shows promise, leaving no trace of mixing association due to the isolation among different chains. 

Regarding all these observations, in our investigation, it was evident that a majority of Bitcoin users opted for centralized mixers as opposed to decentralized alternatives, primarily due to their user-friendliness, intuitive interface, streamlined operations, and the absence of a requirement for advanced technical knowledge.

Besides all these inferences, things are totally different when it comes to Turing-complete-based cryptocurrencies. In such cryptocurrencies like Ethereum, there is no need to setup a mixing framework at a higher layer of the cryptocurrency network, and mixing algorithms can be directly coded and executed in a smart contract, without the need for any centralized intermediary.
This feature gives smart contract developers a wider range of options to create untraceable mixing solutions, resulting in introducing advanced mixing methods like Zero-Knowledge proofs in Tornado Cash \cite{pertsev2019tornadoCash} or ring signatures in Möbius \cite{meiklejohn2018mobius}, all fully decentralized. To the date of this publication, Tornado Cash remains the most popular service of its kind, though its usage has been significantly dropped by more than 95\% after OFAC sanctions enforced \cite{chainalysisCrimeReport, TornadoCashSanctionLink, wang2023zkpMixersEval}. Given these features of Turing-Complete cryptocurrencies, employing centralized mixing frameworks in these networks seems almost pointless.

\section{Possible Attacks on Mixing Services} \label{Sec:Attacks}
Mixing services exist to ensure transaction anonymity, aiming to protect both senders and receivers from being identified at any point in the mixing process. Here we introduce 6 most common attacks within the context of mixing services, which can be performed by a mixer, mixing participants, or any third party who stands outside of the mixing process:


\begin{itemize}
    \item \textbf{Sybil Attack.} A common method to undermine mixing protocols is for an attacker to represent a significant portion of the participants. The larger this deceptive Sybil group, the higher the likelihood that the chosen mixing parties will include the attacker. Essentially, the attacker aims to dominate the system to decrease the anonymity set for others. In pairwise mixing protocols, a Sybil attacker can determine the fund's destination paired with any address. Protocols such as DarkWallet, SharedCoin, and CoinShuffle are vulnerable to cost-free Sybil attacks since they don't charge participants, and creating addresses (or Sybil identities) in Bitcoin and similar virtual currencies is free.

    \item \textbf{DoS Attack.} DarkWallet, SharedCoin, and CoinShuffle don't charge users to join their mix pool, making them vulnerable to denial-of-service (DoS) attacks. An attacker can create multiple identities, each with a unique IP and Bitcoin address, and add them to the mix pool for free. If chosen, the attacker simply doesn't sign the transaction, disrupting the mix. If they create enough identities, they can hinder most transactions without any loss. CoinJoin and its variants, including CoinShuffle, conduct the entire mix in one step and can't impose an upfront fee. Any potential fee would require a different, yet-unproposed protocol. Likewise, Barber et al.’s fair exchange protocol \cite{barber2012fairExchange} doesn't have participation fees, allowing an attacker to disrupt the exchange without any cost.
    
    \item \textbf{Coin Stealing Attack.} This type of attack predominantly occurs through untrusted centralized mixing services that steal user-submitted coins. While some services attempt to mitigate this risk by employing guarantee letters to reassure users about their trustworthiness (like the MixCoin guarantee letter, introduced in chapter 4.1.2), this approach does not provide an absolute guarantee of protection against such attacks.

    \item \textbf{Trojan Attack.} This attack mirrors the Sybil attack, with the attacker actively participating in the mixing process. By transacting with the services, they can trace the mixing pattern and unveil the mixing procedure. Possessing both the sender and receiver addresses allows the attacker to easily discern the transaction pattern. This attack poses the most substantial threat to mixing services as it can expose the entire mixing process, pinpoint mixing transactions, identify related addresses, and uncover those utilizing the mixing services. Several studies have leveraged this attack method to demystify the operations of leading mixing services in the industry \cite{shojaeinasab2022mixing, pakki2021everything, balthasar2017MixerAnalysis2, Novetta}
    
    \item \textbf{Intersection Attack.} The intersection attack focuses on tracking frequent activity by specific wallets either owned by one participant or exhibiting distinct patterns. For instance, in systems like CoinJoin and CoinShuffle, partner peer information is logged on the public blockchain. This allows even non-participating attackers to exploit the vulnerability. In terms of paralleling anonymous communication methods, every attacker in these protocols acts as a global passive observer.
    
    \item \textbf{ML-based Attacks.} In the realm of mixing transactions, a comprehensive and trusted dataset remains elusive. Nonetheless, the WalletExplorer platform offers categorized data for various transaction types, notably for mixing services such as BitcoinFog, Helix, and Bitlaunder\cite{WalletExplorer}. This platform sources its labels predominantly from user submissions and extensive forum analyses, subsequently expanding label classifications based on shared input addresses heuristic deanonymization attacks. In a separate endeavor, Blockstream.info has disseminated a dataset, structured utilizing their proprietary CoinJoin detection methodology and an established method for SharedCoin detection\cite{BitcoinCoinMixingDataSet2023}. The reliability of these datasets, however, remains a subject of debate. Various studies have embarked on experimental pursuits and deep learning-based attacks, yet many confront challenges rooted in the lack of verifiable data.\cite{nan2018bitcoin,wu2021detecting,sun2022lstm}
\end{itemize}

\section{Proposing an Evaluation Framework to Assess Mixing Services} \label{sec:Evaluation}
The paper identifies the most significant functionalities that mixing services utilize to conceal themselves behind blockchain transactions and maintain anonymity. In this section, we introduce an evaluation mechanism, along with specific metrics and criteria, to assess various mixing services and explore their vulnerabilities. To this end, the authors examine various attacks designed to reveal information behind cryptocurrency transactions and undermine anonymity.

In the wake of examining the major challenges posed by attacks on mixing services, we can establish a set of criteria for robust mixing services. Our assessment emphasizes both resilience to attacks and the capacity to obscure cryptocurrency transactions. After introducing various evaluation criteria, Table \ref{tab:Evaluation} presents an assessment of both academic and real-world mixing solutions. The assessment is based on their operational mechanisms, which are discussed in Section \ref{sec:allmixings}.

\begin{itemize}
    \item \textbf{Centralization Concerns:} Centralized tumblers are potential targets for cyber-attacks, legal inquiries, or subpoenas. Compromised records from these services can lead to transaction de-anonymization.
    
    \item \textbf{Unlinkability:} For the highest standard of privacy, it's essential that the connection between the sender and the receiver be entirely severed. A fully unlinkable service ensures that taint analysis or similar techniques can not be traced back to either party, solidifying anonymity.
    Within this context, we can categorize services into 2 different types, Fully Disconnected and Correlated to Pool Size. The second type means that the higher pool size, resulted in a higher anonymization ratio, while in the first type, senders and receivers are completely disconnected from each other.

    \item \textbf{Liquidity:} High liquidity ensures that there are always sufficient funds for mixing, which can prevent delays and improve the anonymity set. Low liquidity mixers might pose the risk of having transactions stalled or linked due to insufficient participants. As the liquidity can be measured only in real-world mixing services, and some of the available data from mixers' liquidities are from several years ago, we mark those services that have enough funds to avoid linking senders to recipients as "High", and others as "Low". 

    \item \textbf{Anonymity Set Size:} The size of an anonymity set is pivotal when assessing the resilience of a mixing service, representing the aggregate of participants or inputs in a specific mixing event. An expansive anonymity set obfuscates the discernment of a coin's flow, while a limited set delineates transaction pathways, potentially undermining user privacy. This metric retains its significance across both traditional and contemporary mixing services. Prominent traditional services like CoinShuffle and CoinJoin fundamentally depend on the integrity of their anonymity sets. In the domain of modern mixers, certain attack vectors have illuminated a paradox. Even as these services generate addresses randomly for each mixing request, there remains a necessity to consolidate funds into a restricted number of addresses to facilitate mixing operations. The inputs tied to these consolidation, or "sweeper," transactions inherently influence the size of the anonymity set. 

    \item \textbf{Log Storing:} Certain centralized mixing services retain mixing logs within their database, which leads to a security risk as sensitive mixing data, such as sender details, recipients, and transaction amounts, could potentially be exposed from the mixing service database. As there is not any reliable data about this metric on services, and nothing can be concluded without accessing to the implementation behind each tumbler, we mark those services who require users to signup before mixing as Signup Required (Signup Req.), since it can be an evidence that they store users' information in their database, which is a security risk. However, it is recommended to use TOR browser for mixing, in order to prevent mixer server from linking the IP addresses to the corresponding crypto addresses.

    \item \textbf{Cost of Mixing (Mixing fee):} The economic viability of using a mixer is crucial for widespread adoption. Services that offer competitive fees while maintaining robust security and privacy features are likely to be preferred. Moreover, an unpredictable fee structure might aid in obfuscation. 
    
    \item \textbf{Auditability:} Some users may prefer services that have undergone third-party audits, ensuring that the underlying code and mechanisms are sound and free of vulnerabilities.

    \item \textbf{Address Reusing:} The redirection of mixed coins to initial addresses or address clusters linked to a user's identity can vitiate the mixing process, thereby reintroducing traceability. Note that this metric is not applicable for smart-contract-based cryptocurrencies, due to their account-based nature.

    \item \textbf{Cross-Chain Mixing:} As the cryptocurrency ecosystem grows, the ability for a mixer to support transactions between different blockchains (cross-chain transactions) can be a valuable feature. It can help users obscure their tracks even further by moving funds between different cryptocurrencies.

    \item \textbf{Sybil / Dos Attack Resistance:} In scenarios where an adversary populates a tumbler with a majority of its own addresses, the integrity of the anonymity set is compromised, thereby facilitating the tracing of coins(Sybil Attack). Moreover, the efficacy of a tumbler can be compromised if a significant portion of its users are colluding or have established identities. In most cases, leveraging mixing fees can be a good obstacle for both Sybil and Dos attacks.
    

    \item \textbf{Coin Stealing Attack Resistance:} When a third-party takes the responsibility to perform mixing, there is a great potential for them to steal the given funds by avoiding sending them the corresponding recipients. This scenario is more likely to happen in terms of centralized services, and several mixers have used mixing guarantee letters, introduced by MixCoin paper \cite{bonneau2014mixcoin}.

    \item \textbf{Trojan / Intersection Attack Resistance:} A mixer's effectiveness is compromised when it uses predictable or discernible transaction mixing patterns. Such patterns, like uniform mixing amounts, and timely patterns, are vulnerable to taint analysis. This vulnerability becomes particularly evident when adversaries deploy Trojan or Intersection attacks, potentially exposing the mixing pattern. Several studies have highlighted successful attacks on prominent Bitcoin services\cite{bissias2014Xim,moser2013mixer1,Novetta,shojaeinasab2022mixing,pakki2021everything}. 
    We assess two important factors in terms of pattern obfuscating: Fee and delay randomization, as discussed in chapter \ref{sec:3-functionalities}. In terms of mixing structure, since the structure of most real-world services has been revealed by prior attacks, we do not discuss them in the table. However, it is vital for a service to be resistant to these types of attacks, and employing dynamic mixing methods can be an innovative way to reach this milestone.

    \item \textbf{Regulatory Compliance:} Depending on the jurisdiction, mixing services might be subject to regulations. Ensuring compliance can prevent potential legal complications for both the service providers and their users. As and example, ShapeShift \cite{ShapeShiftWebsite} employed a Know-Your-Customer policy after being accused of money laundering, which resulted in lower anonymity and more traceability for its users. Due to the lack of reliable data about this metric for most services (except ShapeShift), we don't depict this metric in the table. Nevertheless, Those services that require users to sign up before mixing, are suspected to regulatory compliance. 

    \item \textbf{Front Running Resistance:} Front-running is one of the main challenges in terms of smart-contract-based cryptocurrencies like Ethereum. When talking about mixing, some actions like depositing funds alter the state of the mixing contract, while other actions like withdrawing funds have to rely on the state of the contract to form the cryptographic proofs, which grant them to withdraw their funds. Therefore, if there are multiple concurrent deposit actions issued to the contract, some withdrawing actions will get invalidated by those transactions that modify the state of the contract, and this process can be happened unintentionally by ordinary users, or deliberately by malicious ones \cite{bunz2020zetherFrontRunning1, eskandari2020sokFrontRunning2, le2021AMRmixer}.\\
    Discussing mentioned smart-contract based methods (Möbius, AMR, MixEth, and Tornado Cash), front-running is not applicable in Möbius because of the nature of ring signature scheme \cite{rivest2001ringSig}, not applicable in MixEth due to its shuffling mechanism, and also not cost-efficient in AMR and Tornado Cash \cite{pertsev2019tornadoCash, le2021AMRmixer}.

    
\end{itemize} 

\begin{landscape}
\begin{table}[htbp]
\begin{center}
\vfill
\captionsetup{justification=centering}
\caption{The evaluation of all mixing solutions proposed in academia and market.\\ Rows colors stands as Blue: Decentralized Academic, Orange: Centralized Academic, Green: Decentralized Real-World, and Pink: Centralized Real-World}
\label{tab:Evaluation}
\setlength\tabcolsep{5pt}
\begin{tabular}{l|c:c:c:c:c:c:c:c:c:c:c:c:c:c}
    \toprule
    & \rotatebox{85}{Centralization} & \rotatebox{85}{Unlinkability} & \rotatebox{85}{Liquidity} & \rotatebox{85}{Anonymity Set Size} & \rotatebox{85}{Log Storing} & \rotatebox{85}{Random Delay} & \rotatebox{85}{Random Fee} & \rotatebox{85}{Mixing Fee?} & \rotatebox{85}{Auditable} & \rotatebox{85}{Address Reusing?} & \rotatebox{85}{Cross-Chain Mixing} & 
    \rotatebox{85}{Sybil/Dos Resistance} & \rotatebox{85}{\makecell{Coin Stealing \\ Resistance}} & \rotatebox{85}{\makecell{Front-Running \\ Resistance}}\\
    \midrule
    \rowcolor{lightblue}
    CoinJoin        & \xmark & Cor. Pool Size & - & Limited to Tx Size & \xmark & a. & no fee & \xmark & \cmark & \xmark & \xmark & \xmark & \cmark & n.a.   \\
    \rowcolor{lightblue}
    CoinShuffle     & \xmark & Cor. Pool Size & - & Limited to Tx Size & \xmark      & a. & no fee & \xmark & \cmark & \xmark & \xmark & \xmark & \cmark & n.a. \\
    \rowcolor{lightblue}
    Xim             & \xmark & Fully Disconnected      & - & Participants Count  & \xmark      & a. & \xmark & \cmark & \cmark & \xmark & \xmark & \cmark & \cmark & n.a. \\
    \rowcolor{lightblue}
    CoinParty       & \xmark & Cor. Pool Size & - & Participants Count  & \xmark      & a. & no fee & \xmark & \cmark & \xmark & \xmark & \xmark & \xmark & n.a.    \\
    \rowcolor{lightblue}
    SecureCoin      & \xmark & Cor. Pool Size & - & Participants Count  & \xmark      & a. & a. & \cmark & \cmark & \xmark & \xmark & \cmark & \cmark & n.a.     \\
    \rowcolor{lightblue}
    Möbius          & \xmark & Cor. Pool Size & - & Max 24 users       & \xmark      & a. & a. & \xmark & \cmark & n.a. & \xmark & \cmark & \cmark & \cmark    \\
    \rowcolor{lightblue}
    AMR             & \xmark & Cor. Pool Size & - & Up to Merkle Tree size & \xmark  & a. & a. &  \xmark & \cmark & n.a. & \xmark & \cmark & \cmark & \cmark    \\
    \rowcolor{lightblue}
    MixEth          & \xmark & Cor. Pool Size & - & Participants Count         & \xmark      & a. & a. &  \xmark & \cmark & n.a. & \xmark & \cmark & \cmark & \cmark   \\
    \rowcolor{lightorange}
    Mixcoin         & \cmark & Dep. Implementaion      & - & Participants Count & \xmark       & a. & \cmark & \cmark & \cmark & \xmark & \xmark & \cmark & Guarantee Letter & n.a.   \\
    \rowcolor{lightorange}
    BlindCoin       & \cmark & Dep. Implementaion      & - & Participants Count & \xmark       & a. & \cmark & \cmark & \cmark & \xmark & \xmark & \cmark & Guarantee Letter & n.a.   \\
    \rowcolor{lightorange}
    TumbleBit       & \cmark & Fully Disconnected      & - & Participants Count & \xmark       & a. & a. & \cmark & \cmark & \xmark & \xmark & \cmark & \cmark & n.a.   \\
    \rowcolor{lightorange}
    BSC             & \cmark & Fully Disconnected      & - & Participants Count & \xmark       & a. & a. & \cmark & \cmark & \xmark & \xmark & \cmark & \cmark & n.a.   \\
    \rowcolor{lightorange}
    Obscuro         & \cmark & Dep. Implementaion      & - & Participants Count & \xmark       & a. & a. & \cmark & \cmark & \xmark & \xmark & \cmark & \cmark & n.a.   \\
    \rowcolor{lightgreen}
    BitcoinFog      & \cmark & Cor. Pool Size & High & High Participants & No Signup Req. & - & \cmark & \cmark & \xmark & \xmark & \xmark & \cmark & \xmark & n.a.    \\
    \rowcolor{lightgreen}
    BitLaundry      & \cmark & Cor. Pool Size & Low & Low Participants & No Signup Req. & \xmark & \xmark & \cmark & \xmark & \xmark & \xmark & \cmark & \xmark & n.a.   \\
    \rowcolor{lightgreen}
    Blockchain.com  & \cmark & Cor. Pool Size & High & High Participants & Signup Req.  & - & \xmark & \cmark & \xmark & - & \xmark & \cmark & \xmark & n.a.   \\
    \rowcolor{lightgreen}
    DarkLaunder     & \cmark & Cor. Pool Size & Low & Low Participants & No Signup Req. & \xmark & \xmark & \cmark & \xmark & \cmark & \xmark & \cmark & \xmark & n.a.   \\
    \rowcolor{lightgreen}
    Bitlaunder      & \cmark & Cor. Pool Size & Low & Low Participants & No Signup Req. & \xmark & \xmark & \cmark & \xmark & \cmark & \xmark & \cmark & \xmark & n.a.   \\
    \rowcolor{lightgreen}
    CoinMixer       & \cmark & Cor. Pool Size & Low & Low Participants & No Signup Req. & \xmark & \xmark & \cmark & \xmark & \cmark & \xmark & \cmark & \xmark & n.a.   \\
    \rowcolor{lightgreen}
    Helix           & \cmark & Cor. Pool Size & High & High Participants & No Signup Req. & \cmark & \xmark & \cmark & \xmark & - & \xmark & \cmark & \xmark & n.a.   \\
    \rowcolor{lightgreen}
    Alphabay        & \cmark & Cor. Pool Size & High & High Participants & No Signup Req. & \cmark & \xmark & \cmark & \xmark & - & \xmark & \cmark & \xmark & n.a.   \\
    \rowcolor{lightgreen}
    MixTum          & \cmark & Cor. Pool Size & High & High Participants & No Signup Req. & \cmark & \cmark & \cmark & \xmark & \xmark & \xmark & \cmark & Guarantee Letter & n.a.   \\
    \rowcolor{lightgreen}
    CryptoMixer     & \cmark & Cor. Pool Size & High & High Participants & No Signup Req. & \cmark & \cmark & \cmark & \xmark & \xmark & \xmark & \cmark & Guarantee Letter & n.a.   \\
    \rowcolor{lightgreen}
    Blender         & \cmark & Cor. Pool Size & High & High Participants & No Signup Req. & \cmark & \cmark & \cmark & \xmark & \xmark & \xmark & \cmark & \xmark & n.a.   \\
    \rowcolor{lightgreen}
    ChipMixer       & \cmark & Cor. Pool Size & High & High Participants & No Signup Req. & \cmark & a.     & \xmark & \xmark & \xmark & \xmark & \cmark\textsuperscript{\textbf{1}} & \xmark & n.a.\\
    \rowcolor{lightgreen}
    Wasabi Wallet   & \cmark & Cor. Pool Size & High & Limited to Tx Size & No Signup Req. & - & \xmark & \cmark & \cmark & -     & \xmark & \cmark & \xmark & n.a.  \\
    \rowcolor{lightgreen}
    ShapeShift      & \cmark & Fully Disconnected      & High & n.a.               & Signup Req.    & - & - & \cmark & \xmark & \xmark & \cmark & \cmark & \cmark & n.a.    \\
    \rowcolor{lightgreen}
    Bitmix.bz       & \cmark & Fully Disconnected    & - & High Participants & No Signup Req. & \cmark & \cmark & \cmark & \xmark & \xmark & \xmark & \cmark & Guarantee Letter & n.a.   \\
    \rowcolor{lightgreen}
    Sudoku Wallet   & \cmark & -              & Low & Low Participants & No Signup Req. & \cmark & \cmark & \cmark & \xmark & - & \xmark & \cmark & \cmark & n.a.    \\
    \rowcolor{lightpink}
    Join-Market     & \xmark & Cor. Pool Size & High & Limited to Tx Size & No Signup Req. & a. & a. & \xmark & \cmark & \xmark & \xmark & \cmark & \cmark & n.a.    \\
    \rowcolor{lightpink}
    Tornado Cash    & \xmark & Cor. Pool Size & High & Max $2^{20}$ \textsuperscript{\textbf{*}} & No Signup Req. & a. & \xmark & \xmark & \cmark &  n.a. & \xmark & \cmark & \cmark & \cmark  \\
\bottomrule
\end{tabular}
\end{center}

\begin{tablenotes}
\item[1]{Although mixing fee is donation-based in ChipMixer, its is resilient to these types of attack due to its high liquidity.}\\
\item[*]{This value is corresponded to the maximum size of the Tornado Cash's merkle tree \cite{TornadoCashRepo}.}
\end{tablenotes}

\end{table}
\end{landscape}

As illustrated in the evaluation table \ref{tab:Evaluation}, our analysis is based on pre-defined criteria and offers a comprehensive overview of the current mixing service landscape. Contrary to real-world applications where centralized services dominate, academic research mainly focuses on decentralized solutions. This academic inclination matches user preferences for enhanced privacy and reduced risk from regulatory intervention.

Centralized services come with their own set of inherent risks, including the potential for fraudulent activities and theft of funds, as highlighted in the 'Coin Stealing Resistance' criterion of our evaluation table. On the flip side, most academic solutions are largely immune to such vulnerabilities, with CoinParty being a notable exception.

A significant security vulnerability emerges when a mixing service links multiple peeling chains together through aggregation transactions. In a peeling chain, transactions are divided into smaller parts, which are then distributed to their final destinations. However, as transactions near the end of the peeling chain, the remaining amounts often become too small to be further divided. Some services input these residual amounts into an aggregation transaction, which serves as the starting point for the next peeling chain\cite{shojaeinasab2022mixing}.

By doing so, a traceable link is created between the two separate peeling chains. This linkage could potentially allow an attacker to correlate all transactions related to a particular service, thereby undermining the anonymity of users. In essence, transactions from different chains could be combined into one large cluster, which could be analyzed collectively to identify patterns or isolate individual users.


Another crucial issue revolves around the transparency of the underlying algorithm. Services that disclose their algorithms not only gain more user trust but also invite community contributions. Those solutions using Turing-complete smart contract technology are naturally more credible, due to their auditable and decentralized nature. Comparing early assessments, like the one by Novetta in 2015, with recent reports from CypherTrace and Chainalysis, reveals a noticeable increase in the adoption of mixing services\cite{Novetta,CypherTrace2023,chainalysisCrimeReport}. However, sanctions on services like Tornado Cash have led to a shift towards other centralized alternatives\cite{chainalysisCrimeReport}.

Clearly, each service has its own strengths and drawbacks, and a user who wishes to mix her funds should choose the most appropriate mixer based on her important evaluation metrics. Also, an innovative approach to achieve higher anonymity is to utilize various mixing services in several consecutive rounds, which is referred to as multi-round mixing.

\section{Open Problems and Research Challenges}

The digital frontier of cryptocurrencies, now expanded by the advent of mixing services, marks a significant leap in financial innovation, yet presents unparalleled challenges, particularly in the context of tracing fraudulent transactions and activities. Traditionally, tracking money laundering transactions within a single cryptocurrency ecosystem has been achievable to a certain extent through strategies such as Trojan attacks, ML-based strategies, or Sybil attacks. However, the emergence of cross-chain mixing activities significantly undermines these efforts.

For instance, when dirty money undergoes a series of complex transformations encompassing multiple cryptocurrencies and DeFi solutions, tracing it becomes exponentially difficult. It begins with a Bitcoin mixing service, morphing into Ether via anonymous exchanges, only to be further intertwined with other cryptocurrencies through platforms like Uniswap before it eventually resurfaces as fiat currency. This complex process poses a considerable obstacle in pinpointing the transitional phases where money changes hands and forms.

Moreover, an impending issue surfaces when the money to be laundered originates from fiat currency. The complex network created by combining this phase with cross-chain mixing methods creates an almost untraceable path, amplifying the challenges of monitoring and enforcement. This lack of a fully trackable chain from the path before ramp-on point to ramp-off, combined with the complicated techniques associated with mixing services, constructs a virtually impenetrable shield against current tracking techniques.

In research, a significant issue is the lack of detailed data about mixing services, which makes it difficult to improve machine-learning attack strategies. Without this crucial data, researchers are unable to create and enhance methods that could possibly lead to major breakthroughs in combating money laundering through cryptocurrencies.

Looking forward, a promising avenue of research is to conceptualize a mixing framework that stands resilient against Trojan attacks that are proficient at identifying patterns in transaction flows. Such a framework might hinge on the execution of off-chain transactions, safeguarded by groundbreaking encryption methodologies, zero-knowledge proofs, and Trusted-Execution-Environments, fostering both innovation and security in cryptocurrency transactions.

Additionally, dynamic mixing can serve as an effective solution to prevent Trojan attacks while enhancing the security and anonymity of the mixing process. In essence, dynamic mixing involves using multiple mixing methods in conjunction with each other to make the mixing process look more complicated from the outside, thus hindering attackers from detecting the mixing pattern through their interactions with the service. 

However, working towards this goal is not easy due to the complex nature of the problem, and it demands careful planning to maintain both decentralization and security. Furthermore, embracing the prospect of interdisciplinary research, which merges the strengths of various domains such as cryptography, machine learning, and policy formulation, can create a robust platform to address these open problems. The nexus of these fields could pave the way for innovative solutions that are both secure and preserve the fundamental principles of cryptocurrencies.

In addition to the risks posed by fungible cryptocurrencies, an emerging area for future investigation is the money laundering vulnerabilities associated with non-fungible tokens (NFTs). NFTs combine the pseudonymity and international accessibility of crypto-assets with the subjective pricing systems often seen in the fine art market\cite{mosna2023nfts}. This blend of characteristics creates a complex challenge for anti-money laundering frameworks, as the highly variable and subjective pricing of NFTs provides opportunities for value manipulation, making them particularly susceptible to illicit financial activities.

Conclusively, as the landscape of cryptocurrencies evolves, it requires a proactive approach in addressing the imminent challenges and fostering an environment that nurtures innovation while upholding the principles of security and privacy. Finding an optimal pathway that facilitates advancements in technology while maintaining strict regulations poses a formidable challenge, yet stands as a vital undertaking for the community in the coming years.

\section{Conclusions}

In this comprehensive study, the authors have introduced a novel evaluation framework for blockchain-based mixing services, marking the first instance where such criteria have been systematically defined and applied. The manuscript provides an in-depth analysis that bridges academic theories and real-world applications, making it both a tutorial for beginners and a resource for experts to identify the strengths and weaknesses of each solution, as well as potential attack vectors. This study has rigorously analyzed 31 mixing services implementations and has gone beyond theoretical discussions by conducting real-world attacks to reveal the operational mechanisms of existing black-box mixing services.

A key finding highlights the prevalent use of a conventional set of techniques—such as aggregation, peeling chains, and the integration of random delays and fees—in current real-world mixing services. These findings contrast sharply with the innovative and diverse methods that have been theorized in academic circles.

Regulatory compliance and potential legal implications are particularly relevant for centralized mixing services, casting doubt on their long-term sustainability. Additionally, the absence of dynamic mixing strategies, which would combine multiple techniques, represents a significant shortfall in augmenting both the security and anonymity aspects of blockchain transactions.

Furthermore, the manuscript underscores the need for mixing services that are compatible across multiple blockchain platforms, a challenge that also presents a significant opportunity for future innovation. Overcoming the technical barriers to achieving cross-chain mixing would catalyze further advancements in this rapidly growing domain.

Consequently, the study emphasizes the need for closer integration between academic research and industry practices to nurture a blockchain ecosystem that is not only secure but also robustly anonymous.

\section*{Acknowledgments}
We would like to acknowledge the financial support of Mastercard Co. and Mathematics of Information Technology and Complex Systems (MITACS) under IT34028 Mitacs Accelerate. 

\section*{CRediT authorship contribution statement}
\textbf{Alireza Arbabi} Conceptualization, Formal analysis, Investigation, Visualization, Writing - Original Draft, Writing - Review \& Editing.
\textbf{Ardeshir Shojaeinasab} Conceptualization, Methodology, Validation, Writing - Original Draft, Writing - Review \& Editing, Supervision, Project administration. \textbf{Behnam Bahrak} Writing - Review \& Editing, Supervision. \textbf{Homayoun Najjaran} Writing - Review \& Editing, Supervision.

\bibliographystyle{unsrt}  
\bibliography{main}

\end{document}